\documentclass{aa}  

\usepackage{graphicx}
\usepackage{color}
\usepackage{amsmath}
\usepackage{natbib}
\usepackage{txfonts}
\usepackage{hyperref}
\hypersetup{breaklinks=true, colorlinks=true, citecolor=blue, 
linkcolor=black, urlcolor=blue}
\begin{document}

   \title{PKS\,1502+106: A high-redshift $\textrm{\textit{Fermi}}$ blazar at
extreme angular resolution}

   \subtitle{Structural dynamics with VLBI imaging up to 86\,GHz}

   \author{V. Karamanavis\inst{1}\thanks{To whom correspondence
should be addressed.}
          \and
          L. Fuhrmann\inst{1}
          \and
          T. P. Krichbaum\inst{1}
          \and
          E. Angelakis\inst{1}
          \and
          J. Hodgson\inst{1}
          \and
          I. Nestoras\inst{1}
          \and
          I. Myserlis\inst{1}
          \and
          J.\,A.\,Zensus\inst{1}
          \and
          A. Sievers\inst{2}
          \and
          S. Ciprini\inst{3,4,5}
          }

   \institute{Max-Planck-Institut f\"{u}r Radioastronomie, Auf dem H\"{u}gel
69, D-53121 Bonn, Germany\\
\email{\href{mailto:vkaraman@mpifr-bonn.mpg.de}{vkaraman@mpifr-bonn.mpg.de}}
         \and
         Instituto de Radio Astronom\'{i}a Milim\'{e}trica, Avenida Divina
Pastora 7, Local 20, E-18012, Granada, Spain
         \and
         Agenzia Spaziale Italiana (ASI) Science Data Center, I-00133 Roma,
Italy
         \and
         Istituto Nazionale di Astrofisica -- Osservatorio Astronomico di Roma,
I-00040 Monte Porzio Catone (Roma), Italy
         \and
         Istituto Nazionale di Fisica Nucleare, Sezione di Perugia,
I-06123 Perugia, Italy
         }

   \date{Received Aug 21, 2015; accepted Oct 29, 2015}


  \abstract
  {Blazars are among the most energetic objects in the Universe. In
2008 August, \textit{Fermi}/LAT detected the blazar PKS\,1502+106 showing a
rapid and strong $\gamma$-ray outburst followed by high and variable flux over
the next months. This activity at high energies triggered an intensive
multi-wavelength campaign covering also the radio, optical, UV, and X-ray bands
indicating that the flare was accompanied by a simultaneous outburst at
optical/UV/X-rays and a delayed outburst at radio bands.}
   {In the current work we explore the phenomenology and physical
conditions within the ultra-relativistic jet of the $\gamma$-ray blazar
PKS\,1502+106. Additionally, we address the question of the spatial
localization of the MeV/GeV-emitting region of the source.}
   {We utilize ultra-high angular resolution mm-VLBI observations at 43 and
86\,GHz complemented by VLBI observations at 15\,GHz. We also employ
single-dish radio data from the F-GAMMA program at frequencies matching the VLBI
monitoring.}
   {PKS\,1502+106 shows a compact core-jet morphology and fast superluminal
motion with apparent speeds in the range 5--22\,c. Estimation of Doppler
factors along the jet yield values between ${\sim} 7$ up to ${\sim} 50$.
This Doppler factor gradient implies an accelerating jet. The
viewing angle towards the source differs between the inner and outer jet, with
the former at $\theta \sim 3^{\circ}$ and the latter at $\theta \sim 1^{\circ}$,
after the jet bends towards the observer beyond 1\,mas.
The de-projected opening angle of the ultra-fast, magnetically-dominated jet is
found to be $(3.8 \pm 0.5)^{\circ}$. A single jet component can be
associated with the pronounced flare both at high-energies and in radio bands.
Finally, the $\gamma$-ray emission region is localized at ${\leq}5.9$\,pc away
from the jet base.}
   {}


   \maketitle
%
\section{Introduction}

Radio-loud active galactic nuclei
(AGN) and blazars in particular -- constituting their beamed population -- are
among the most copious and variable emitters of radiation in the Universe. They
intrinsically feature double, opposite-directed fast plasma outflows. Those are
most likely driven by the conversion of gravitational into
electromagnetic and kinetic energy, in the immediate vicinity of spinning
supermassive black holes (SMBHs) \citep{1977MNRAS.179..433B} or from the
accretion disk surrounding them \cite[][for reviews cf.
\citealt{2001Sci...291...84M,2001PThPS.143..182B,2003NewAR..47..667M}]{
1982MNRAS.199..883B}. The linear extent of those jets greatly
exceeds the dimensions of their hosts and can reach undisrupted, up to a few
million parsecs (pc).

Their extreme phenomenology includes fast superluminal motion at parsec and
sub-parsec scales, high degree of radio and optical linear polarization, along
with rapid broad band flux density and polarization variability at time scales
down to minutes. Their double-humped spectral energy distributions (SEDs), first
peak between radio and soft X-rays, due to synchrotron emission and then at keV
to GeV energies, likely owing to inverse Compton (IC) scattering
\citep[see e.g.][]{1997ARA&A..35..445U, 1999APh....11..159U}. However, the
detailed processes giving rise to blazar characteristics are still under intense
debate. Outstanding questions -- among others -- include:
(i) which mechanism(s) drive their rapid variability across the whole
electromagnetic spectrum, (ii) where in the jet does the high-energy emission
originate and (iii) what is the target photon field for IC up-scattering; the
dusty torus and ambient infrared (IR) photon field, broad-line region (BLR) or
accretion disk photons? Through their observational predictions, different
scenarios can be put to the test. Theoretical considerations indicate that the
bulk of high-energy $\gamma$-ray emission should be emanating
from regions of low ambient density otherwise the pair production process
$\gamma + \gamma \rightarrow e^{-} + e^{+}$, would prohibit $\gamma$-ray photons
to escape the MeV/GeV production region, introducing a high $\gamma$-ray opacity
environment \citep{2012ApJ...755..147D, 2012arXiv1209.2291T}. Nonetheless,
observational evidence is still conflicting, indicating that GeV-emission can be
produced close to the central engine, within the broad-line region (BLR), or
even in the vicinity of the accretion disk some $100$ Schwarzschild radii from
the SMBH \citep[e.g.][]{1995ApJ...441...79B}. Other findings lend support to the
``larger-distance scenario'', whereby MeV/GeV photons originate from shock-shock
interaction or turbulent cells within the plasma flow \citep[see
e.g.][]{1995A&A...297L..13V, 2014ApJ...780...87M}.

Correlated variability and flare timing analyses, between radio and $\gamma$-ray
bands place the production site of $\gamma$ rays few pc
upstream of the mm-band unit-opacity surface \citep{2014MNRAS.441.1899F},  while
other findings point towards traveling or standing shocked regions downstream
the jet  \citep{1995A&A...297L..13V}. With the bulk of blazar activity taking
place in regions close to the central engine, very-long-baseline interferometry
(VLBI) at mm wavelengths is essential for addressing those questions. This
technique is able to deliver both the highest angular resolution attainable and
penetrate the opacity barrier which renders those regions inaccessible to
lower frequency observations. As an example, work based on 7\,mm VLBI and flux
monitoring of OJ\,287 places the $\gamma$-active region more than 14\,pc away
from its central engine \citep{2011ApJ...726L..13A}.

In 2008 August, the \textit{Fermi}-GST \citep[Gamma-ray Space
Telescope, hereafter \textit{Fermi},][]{2007AIPC..921....3R,
2009ApJ...697.1071A} detected the blazar PKS\,1502+106 showing a rapid and
strong $\gamma$-ray outburst followed by high and variable flux over the next
months \citep{2008ATel.1650....1C, 2010ApJ...710..810A}. This activity at high
energies triggered an intensive multi-wavelength campaign, covering the radio,
optical, UV, and X-ray bands. The outburst was accompanied by a simultaneous
flare at optical/UV/X-rays \citep{2011A&A...526A.125P},
with a significantly delayed counterpart at radio bands
\citep{2014MNRAS.441.1899F}.

PKS\,1502+106 (OR\,103, S3\,1502+10) at a redshift of $z = 1.8385$, $D_{\rm L}
= 14176.8$ Mpc \citep{2008ApJS..175..297A}, is a powerful blazar classified as a
flat spectrum radio quasar (FSRQ) with the mass of its central engine
${\sim} 10^{9}M_{\odot}$ \citep[][and references
therein]{2010ApJ...710..810A}.
In X-rays, it is known as a significantly variable source
\citep{1994ApJ...436L..59G}. Radio interferometric observations with the VLA at
1.4\,GHz reveal the large-scale morphology of PKS\,1502+106, with a straight jet
at a position angle (PA) of about $-160^{\circ}$ \citep{2007ApJS..171..376C}.
Previous VLBI findings suggest a curved, asymmetric jet and a
multi-component, core-dominated source \citep{1996ApJ...468..543F} with very
fast apparent superluminal motion of up to $(37 \pm 9)\,c$
\citep{2004A&A...421..839A}.

In the following, we present a comprehensive VLBI study of PKS\,1502+106 at
three frequencies, namely 15, 43, and 86\,GHz (wavelengths of 20, 7, and 3 mm,
respectively). Our millimeter-VLBI monitoring of PKS\,1502+106 was triggered by
the flaring activity that the source underwent and started in early
2008. Data at 43 and 86\,GHz span the time period between 2009 May and 2012
May comprising six observing epochs. VLBI monitoring at mm wavelengths
is complemented by nineteen epochs of VLBA observations at 15\,GHz from the
MOJAVE program \citep{2009AJ....138.1874L}.

Throughout this paper we adopt $S \propto \nu^{+ \alpha}$, where $S$ is the
radio flux density, $\nu$ the observing frequency, and $\alpha$ the spectral
index, along with the following cosmological parameters $H_{0} = 71\,{\rm
km\,s}^{-1}\,{\rm Mpc}^{-1}$, $\Omega_{\rm m} = 0.27$, $\Omega_{\Lambda} =
0.73$.

At the redshift of PKS\,1502+106 an angular separation of 1 milliarcsecond (mas)
translates into a linear distance of 8.53\,pc and a proper motion of
1\,mas\,yr$^{-1}$ corresponds to an apparent superluminal speed of 79\,$c$.

The paper is structured as follows. In Section 2 an introduction to
the technical aspects of the observations and the data reduction techniques
is presented. In Sect. 3 we present our findings concerning the phenomenology
of PKS\,1502+106, while in Sect. 4 the deduced physical parameters are shown.
Sect. 5 discusses our findings in a broader context and in conjunction
with previous work by other authors. In Sect. 6 a summary and our
conclusions are presented.

%
\section{Observations and data reduction}

\subsection{High frequency 43\,GHz and 86\,GHz GMVA data}\label{sect:mmVLBI_obs}

The mm-VLBI observations with the Global Millimeter VLBI Array
(GMVA\footnote{\texttt{www.mpifr-bonn.mpg.de/div/vlbi/globalmm/}})
comprise 6 epochs spanning the period between 2009 May 7 and 2012 May 18 with
observations obtained twice per year, apart from 2010 October. The experiments
included the following stations; in Europe: Effelsberg(100\,m, Ef), Onsala
(20\,m, On), Pico Veleta (30\,m, Pv), Plateau de Bure (6$\,\times\,$15\,m phased
array, Pb), Mets\"{a}hovi (14\,m, Mh), Yebes (40\,m, Ys); in the USA, the Very
Long Baseline Array (VLBA) stations that are equipped with 86\,GHz receiving
systems (8$\,\times\,$25\,m).
For a complete presentation of the observational setup refer to Tables
\ref{table:2} and \ref{table:3}. The GMVA experiments were carried out in dual
polarization mode at the stations where this setup is available (Ys and On
record only LCP) at a bit rate of 512\,Mbits/s. The observing strategy at
86\,GHz comprised seven-minute scans of PKS\,1502+106 bracketed by scans of
calibrators for the full duration of the session. Calibrator sources
included 3C\,273, 3C\,279, 3C\,454.3, and PKS\,1510-089. The 43\,GHz scans,
using only the VLBA, were obtained in gaps when European stations were
performing pointing calibration. Data were correlated at the MK4 correlator of
the MPIfR in Bonn, Germany.

\subsubsection{\textit{A priori} calibration}

After correlation, the raw data were loaded to, and inspected with the National
Radio Astronomy Observatory's (NRAO) Astronomical Image processing System (AIPS)
software \citep{1990apaa.conf..125G}. Phase calibration, being the
most crucial step due to atmospheric and instrumental instabilities, was done
using the task \texttt{FRING} within AIPS in the following manner. First, an
initial manual phase calibration was applied incrementally to the data
\citep[][]{2012A&A...542A.107M}. That is,
calculating single subband phases and delays, first for the VLBA
antennas and then for European stations, and applying them to the data. These
phases and delays were estimated with respect to a reference
antenna using a set of visibility data from a bright source, among those
listed in Sect. \ref{sect:mmVLBI_obs}. Subsequently, global fringe fitting was
performed and phases, delays, and rates were deduced and applied to the
multi-band data. We performed the initial \textit{a priori} amplitude
calibration within AIPS and applied an atmospheric opacity correction to the
data. The visibility amplitudes were calibrated using the measured system
temperatures and telescope gain curves. Finally, fully calibrated visibility
data were exported to DIFMAP for further analysis.

\subsubsection{Imaging and model fitting}\label{red:Difmap}

For the subsequent imaging and analysis we have used the 
DIFMAP software \citep{1997ASPC..125...77S, 1994BAAS...26..987S}. After
importing the visibility data, continuous deconvolution cycles were performed
with the application of the \texttt{CLEAN} algorithm
\citep{1974A&AS...15..417H}, followed by phase self-calibration. Amplitude
self-calibration, with decreasing integration time, was performed after every
sequence of \texttt{CLEAN} and phase self-calibration and only when a
representative model for the source had been obtained. Extreme caution was
exercised in order not to artificially and/or irreversibly alter the morphology
of the source, by introducing or eliminating spurious/fake or real source
features. This was accomplished by comparing the model with the visibilities and
by preventing the standard deviation of visibility amplitudes to vary by more
than 10\%.

In order to parameterize each image and quantitatively handle the relevant
quantities, the \texttt{MODELFIT} algorithm within DIFMAP was employed.
To simplify the analysis and reduce the number of unconstrained parameters we
only fit circular Gaussian components. The fit parameters are: the component
flux density, its size expressed as the full width at half maximum (FWHM) of the
two-dimensional Gaussian and its position angle (PA) expressed in degrees 
(between $0^{\circ}$ and $180^{\circ}$ in the clockwise and
$0^{\circ}$ to $-180^{\circ}$ in the counterclockwise direction). We have
modeled the data at each frequency and each observing epoch separately.
Initially the number of components is unrestricted. The limit to the number of
components was set by testing whether the addition of a new one, causes 
the $\chi^{2}$ value for the fit to decrease significantly.

Uncertainties of model fitting are taken as follows. For the component flux
density an uncertainty of 10$\%$ is a reasonable estimate that is widely adopted
\citep{2009AJ....138.1874L, 2013AJ....146..120L}.
The uncertainty in the position is set to $1/5$ of the beam size, if the
component is unresolved, otherwise we take $1/5$ of the component FWHM as an
estimate of the uncertainty. This is increased to half a beam for unresolved
knots at 86\,GHz. The beam size for each observing epoch is calculated through
$b_{\phi} = \sqrt{b_{\rm maj} ~ b_{\rm min}}$ \citep{2005astro.ph..3225L}.
Uncertainty in the PA is the direct translation of the positional
uncertainty ($\Delta {\rm X}$) from mas to degrees using the simple
trigonometric formula $\Delta {\rm PA} = \arctan \left( \Delta {\rm X} / r
\right)$, with $r$ the radial separation in mas.

To define robust jet components we adopt the following criterion. A
physical region of enhanced emission within the jet should repeatedly appear
in three consecutive observing epochs. We relax this criterion for 86\,GHz data
and only when images in specific intermediate epochs suffer from low dynamic
range, preventing us from discerning a component whose presence is indicated by
data in adjacent epochs. In the process of obtaining the final kinematical model
at each frequency, the component separation, size, and flux density were used as
parameters combined with the requirement that they should vary smoothly, in a
way compatible with physical processes.

\subsection{15\,GHz MOJAVE survey data}

In this work we have also made use of the publicly available data from the
Monitoring Of Jets in Active galactic nuclei with VLBA Experiments
(MOJAVE\footnote{\texttt{www.physics.purdue.edu/MOJAVE}}) program
\citep{2009AJ....138.1874L}. Data on PKS\,1502+106 comprise 19 observing epochs
that span the period between 2002 August 12 and 2011 August 15. Apart from
few occasions, all ten VLBA stations participate in the observations (see Table
\ref{table:1}). Data are provided as fully self-calibrated visibilities and as
such, imaging and model fitting were performed, following the steps described in
Sect. \ref{red:Difmap}.

\subsection{Single-dish 15, 43, and 86\,GHz F-GAMMA data}

Single-dish, multi-frequency monitoring data were obtained within the
framework of the long-term \textit{Fermi}-GST AGN Multi-frequency Monitoring
Alliance
(F-GAMMA\footnote{\texttt{www.mpifr-bonn.mpg.de/div/vlbi/fgamma/fgamma.html}})
program \citep{2007AIPC..921..249F, 2010arXiv1006.5610A}. F-GAMMA is the
coordinated effort for simultaneous observations at 11 bands in the range
between 2.64\,GHz and 345\,GHz. Regular monthly observations including the
Effelsberg 100-m and IRAM 30-m telescopes are closely coordinated to ensure
maximum coherency. The observing strategy utilizes cross-scans; that is slewing
over the source position, both in azimuth and elevation. After initial
flagging, pointing offset, opacity, sensitivity, and gain-curve corrections are
applied to the data \citep[for details see][]{2014MNRAS.441.1899F,
2015A&A...575A..55A, 2015Nestoras}.

From the broad F-GAMMA frequency coverage we use only the bands matching our
mm-VLBI frequencies, namely at 14.60, 43.00 and 86.24\,GHz. A detailed
single-dish, multi-wavelength study using the full spectral coverage of F-GAMMA,
as well as optical/X-ray data will appear in a forthcoming article.

\subsection{$\textrm{\textit{Fermi}}$/LAT $\gamma$-ray data}

\textit{Fermi} is the latest generation space-bound telescope built for, and
dedicated to detailed observations of the high-energy sky. Since its launch in
2008 its two instruments the Large Area Telescope
\citep[LAT,][]{2009ApJ...697.1071A} and the Gamma-ray Burst Monitor (GBM) scan
the full sky about once every 3 hours in the energy range from
${\sim} 10$\,keV to ${>}300$\,GeV providing unprecedented temporal and spectral
coverage.

Here, we employ the monthly-binned LAT $\gamma$-ray light curve of
PKS\,1502+106, already presented in \cite{2014MNRAS.441.1899F}, covering
the time period between MJD\,$54707$ (2008.66, 2008 Aug. 29) and MJD\,$55911$
(2011.96, 2011 Dec. 16). Details on the $\gamma$-ray flux ($E>100$\,MeV)
light curve and analysis are given in the aforementioned reference \citep[see
also][]{2010ApJ...710..810A}.

\begin{table*}
\caption{Summary of the six observing epochs of PKS $1502{+}106$ at $43$\,GHz
observed during GMVA sessions with the VLBA.}  
\label{table:2}      
\centering
\footnotesize
\begin{tabular}{l l c c c c c r}
\hline\hline       
Obs. Date         & Array elements       & S$_{\rm peak}^{(1)}$ & rms$^{(2)}$& S$_{\rm total}^{(3)}$ & $b_{\rm maj}^{(4)}$ & $b_{\rm min}^{(5)}$ & PA$^{(6)}$\\
                  &		         & $( {\rm Jy\,beam}^{-1})$ & $( {\rm mJy\,beam}^{-1})$ & (Jy) & (mas) & (mas) & ($^{\circ}$) \\
\hline\noalign{\medskip}
2009-05-08$^{\textrm{I}}$   & VLBA$_{8}$ & $3.473$  & $0.92$ & $3.697$ &  $0.56$  &  $0.17$  &  $-21.0$  \\
2009-10-13$^{\textrm{II}}$  & VLBA$_{8}$ & $2.199$  & $1.15$ & $2.562$ &  $0.73$  &  $0.17$  &  $-17.7$  \\
2010-05-07$^{\textrm{III}}$ & VLBA$_{8}$ & $0.533$  & $0.67$ & $0.647$ &  $0.60$  &  $0.18$  &  $-20.7$  \\
2011-05-08$^{\textrm{IV}}$  & VLBA$_{8}$ & $0.926$  & $0.91$ & $1.179$ &  $0.61$  &  $0.24$  &  $-20.2$  \\
2011-10-09$^{\textrm{V}}$   & VLBA$_{8}$ & $0.953$  & $0.72$ & $1.130$ &  $0.58$  &  $0.18$  &  $-21.8$  \\
2012-05-18$^{\textrm{VI}}$  & VLBA$_{8}$ & $0.821$  & $0.79$ & $0.919$ &  $0.61$  &  $0.20$  &  $-18.2$  \\ 
\noalign{\smallskip}
\hline                  
\end{tabular}
\tablefoot{Station designations:
VLBA$_{8}=$ Br -- Fd -- Kp -- La -- Mk -- Nl -- Ov -- Pt;
$^{(1)}$Peak flux density, $^{(2)}$off-source rms noise, and $^{(3)}$ total
flux density of the \texttt{MODELFIT} image. $^{(4)}$Major and $^{(5)}$minor
axes along with the $^{(6)}$position angle of the restoring beam.
Parameters listed above correspond to the final, uniform weighted, untapered
image. Labels \textrm{I} through \textrm{VI} correspond to the
quasi-simultaneous data forming the spectra in Fig. \ref{compSpectraAll}
(see Sect. \ref{sect:spectra}).}
\end{table*}

\begin{table*}
\caption{Summary of the six observing epochs of PKS $1502{+}106$ at $86$\,GHz
observed with the GMVA.}  
\label{table:3}      
\centering
\footnotesize
\begin{tabular}{l l c c c c c r}
\hline\hline       
Obs. Date  & Array elements & S$_{\rm peak}^{(1)}$ & rms$^{(2)}$ & S$_{\rm total}^{(3)}$ & $b_{\rm maj}^{(4)}$ & $b_{\rm min}^{(5)}$ & PA$^{(6)}$\\
           &                & $( {\rm Jy\,beam}^{-1})$ & $( {\rm mJy\,beam}^{-1})$ & (Jy) & (mas) & (mas) & ($^{\circ}$) \\
\hline\noalign{\medskip}
 2009-05-07$^{\textrm{I}}$   &   VLBA$_{8}$ + Pb + Pv + On + Ef              &  $2.230$    &  $0.61$  & $2.853$ & $0.27$  &  $0.06$  &  $-3.9 $  \\
 2009-10-13$^{\textrm{II}}$  &   VLBA$_{8}$ + Pv + Pb + On + Ef              &  $0.878$    &  $0.91$  & $1.101$ & $0.20$  &  $0.05$  &  $-5.2 $  \\
 2010-05-06$^{\textrm{III}}$ &   VLBA$_{7}$ ($-$Nl) + Pv + Pb + On + Mh + Ef &  $0.487$	   &  $0.36$  & $0.614$ & $0.32$  &  $0.08$  &  $-10.4$  \\
 2011-05-07$^{\textrm{IV}}$  &   VLBA$_{8}$ + Pv + Pb + On + Mh + Ef	     &  $0.762$    &  $0.53$  & $0.875$ & $0.29$  &  $0.07$  &  $-4.8 $  \\
 2011-10-09$^{\textrm{V}}$   &   VLBA$_{8}$ + Ys + Pv + On + Ef              &  $0.571$    &  $1.67$  & $0.652$ & $0.20$  &  $0.07$  &  $-4.2 $  \\
 2012-05-17$^{\textrm{VI}}$  &   VLBA$_{7}$ ($-$Fd) + Ef + On + Pv + Ys      &  $0.535$    &  $1.18$  & $0.616$ & $0.24$  &  $0.06$  &  $-5.8 $  \\
\noalign{\smallskip}
\hline                  
\end{tabular}
\tablefoot{Station designations:
VLBA$_{8}$ is same as above;
Ef -- Effelsberg;
Mh -- Mets\"{a}hovi;
On -- Onsala;
Pb -- Plateau de Bure (phased array);
Pv -- Pico Veleta;
Ys -- Yebes;
$^{(1)}$Peak flux density, $^{(2)}$off-source rms noise, and $^{(3)}$ total
flux density of the \texttt{MODELFIT} image. $^{(4)}$Major and $^{(5)}$minor
axes along with the $^{(6)}$position angle of the restoring beam.
Parameters listed above correspond to the final, uniform weighted, untapered
image. Labels \textrm{I}--\textrm{VI} correspond to the
quasi-simultaneous data forming the spectra in Fig. \ref{compSpectraAll}
(see Sect. \ref{sect:spectra}).}
\end{table*}

\begin{table*}
\caption{Summary of the nineteen VLBA observing epochs at $15$\,GHz of PKS
1502{+}106 obtained within the MOJAVE monitoring program.}    
\label{table:1}      
\centering
\footnotesize
\begin{tabular}{l l c c c c c r}
\hline
\hline       
Obs. Date & Array elements & S$_{\rm peak}^{(1)}$ & rms$^{(2)}$ & S$_{\rm total}^{(3)}$ & $b_{\rm maj}^{(4)}$ & $b_{\rm min}^{(5)}$ & PA$^{(6)}$\\
          &                & $( {\rm Jy\,beam}^{-1})$ & $( {\rm mJy\,beam}^{-1})$ & (Jy) & (mas) & (mas) & ($^{\circ}$) \\
\hline\noalign{\medskip}
 2002-08-12                  &   VLBA$_{10}$      & $1.169$ & $1.52$ & $1.615$ & $0.97$ & $0.44$ & $-2.8 $ \\	
 2003-03-29                  &   VLBA$_{10}$      & $1.288$ & $1.49$ & $1.180$ & $0.95$ & $0.43$ & $ 1.4 $ \\	
 2004-10-18                  &   VLBA$_{10}$      & $0.561$ & $0.97$ & $0.995$ & $1.05$ & $0.42$ & $-8.0 $ \\	
 2005-05-13                  &   VLBA$_{10}$      & $0.509$ & $0.85$ & $0.867$ & $1.04$ & $0.46$ & $-8.2 $ \\	
 2005-09-23                  &   VLBA$_{10}-$Nl   & $0.772$ & $0.58$ & $1.119$ & $0.93$ & $0.41$ & $-0.7 $ \\	
 2005-10-29                  &   VLBA$_{10}$      & $0.848$ & $0.53$ & $1.144$ & $1.03$ & $0.46$ & $-8.8 $ \\	
 2005-11-17                  &   VLBA$_{10}-$Br   & $0.897$ & $1.02$ & $1.176$ & $1.21$ & $0.46$ & $-11.8$ \\	
 2006-07-07                  &   VLBA$_{10}$      & $1.164$ & $1.81$ & $1.527$ & $1.09$ & $0.45$ & $-9.6 $ \\	
 2007-08-16                  &   VLBA$_{10}$      & $1.099$ & $0.83$ & $1.509$ & $0.97$ & $0.49$ & $-7.8 $ \\	
 2008-06-25                  &   VLBA$_{10}-$Br   & $1.328$ & $0.77$ & $1.742$ & $1.10$ & $0.40$ & $-7.7 $ \\	
 2008-08-06                  &   VLBA$_{10}$      & $1.327$ & $0.75$ & $1.711$ & $1.04$ & $0.48$ & $-7.3 $ \\	
 2008-11-19                  &   VLBA$_{10}$      & $1.577$ & $0.68$ & $1.984$ & $1.04$ & $0.50$ & $-6.4 $ \\	
 2009-03-25$^{\textrm{I}}$   &   VLBA$_{10}-$Hn   & $2.791$ & $0.93$ & $3.165$ & $1.12$ & $0.50$ & $-3.2 $ \\	
 2009-12-10$^{\textrm{II}}$  &   VLBA$_{10}$      & $1.191$ & $0.77$ & $1.480$ & $1.02$ & $0.47$ & $-12.7$ \\	
 2010-06-19$^{\textrm{III}}$ &   VLBA$_{10}$      & $0.712$ & $0.58$ & $0.981$ & $1.14$ & $0.47$ & $-12.8$ \\	
 2010-08-27                  &   VLBA$_{10}$      & $0.735$ & $0.51$ & $0.991$ & $0.99$ & $0.44$ & $-5.7 $ \\	
 2010-11-13                  &   VLBA$_{10}$      & $0.860$ & $0.59$ & $1.116$ & $1.05$ & $0.42$ & $-9.4 $ \\	
 2011-02-27$^{\textrm{IV}}$  &   VLBA$_{10}$      & $0.978$ & $0.70$ & $1.231$ & $0.93$ & $0.42$ & $ 0.5 $ \\	
 2011-08-15$^{\textrm{V}}$   &   VLBA$_{10}$      & $1.106$ & $0.91$ & $1.350$ & $1.17$ & $0.44$ & $-1.1 $ \\	
 \noalign{\smallskip}
 \hline                  
\end{tabular}
\tablefoot{VLBA$_{10}$ station designations:
Br -- Brewster; 
Fd -- Fort Davis;
Hn -- Hancock;
Kp -- Kitt Peak;
La -- Los Alamos;
Mk -- Mauna Kea;
Nl -- North Liberty;
Ov -- Owens Valley; 
Pt -- Pie Town;
Sc -- St. Croix.
$^{(1)}$Peak flux density, $^{(2)}$off-source rms noise, and $^{(3)}$ total
flux density of the \texttt{MODELFIT} image. $^{(4)}$Major and $^{(5)}$minor
axes along with the $^{(6)}$position angle of the restoring beam.
Parameters listed above correspond to the final, uniform weighted, untapered
image.Labels \textrm{I}--\textrm{VI} correspond to the quasi-simultaneous data
forming the spectra in Fig. \ref{compSpectraAll} (see
Sect. \ref{sect:spectra}).}
\end{table*}

%
\section{PKS\,1502+106 jet phenomenology}\label{Sect:Phenomenology}

\subsection{Parsec-scale jet morphology}\label{pcScale}

\begin{figure*}
   \centering
\includegraphics[width=\textwidth]{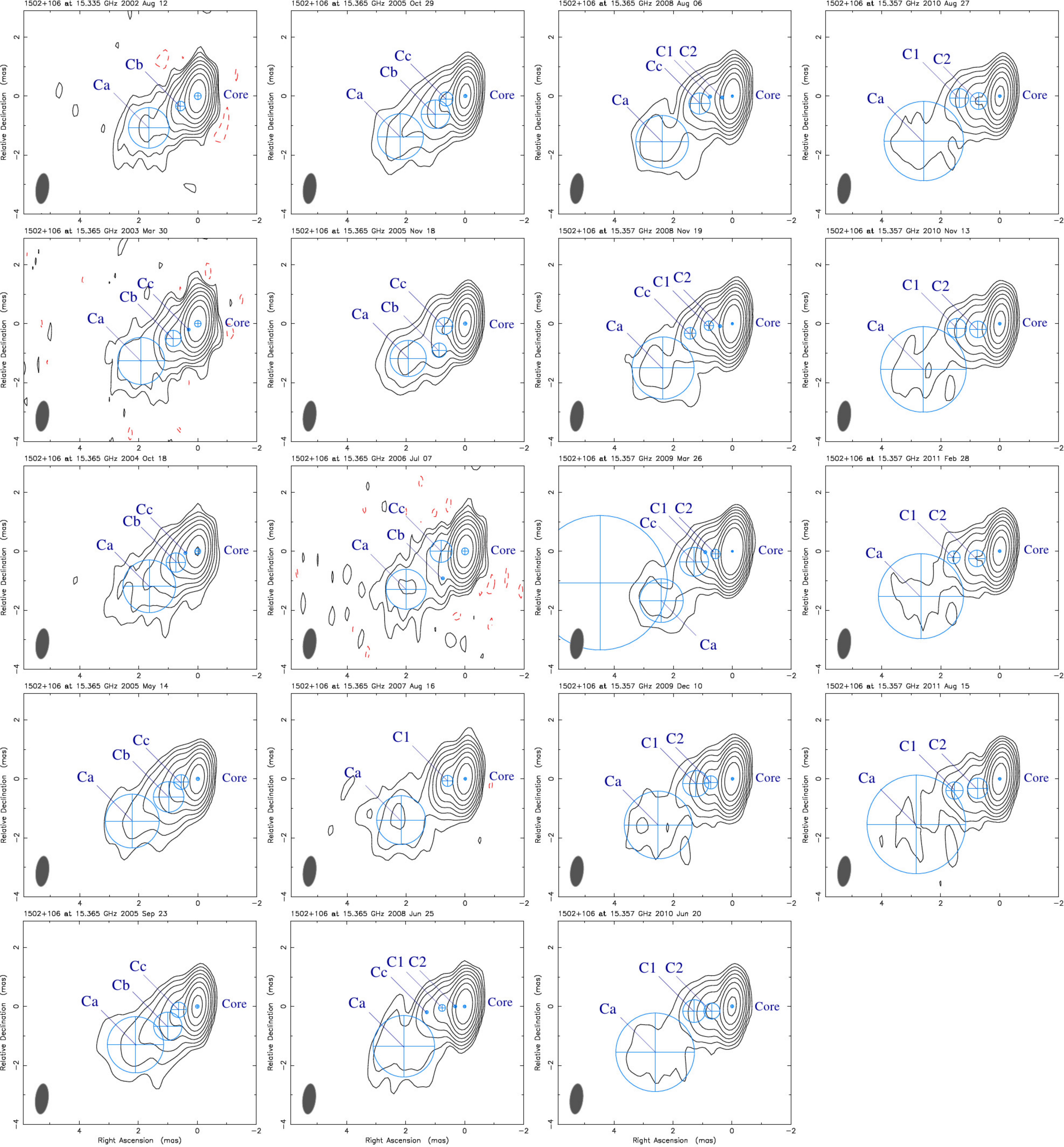}
\caption{Uniformly weighted, untapered \texttt{MODELFIT} images of PKS\,1502+106
at $15$\,GHz. Contour levels correspond to $-0.15\%$, $0.15\%$, $0.3\%$,
$0.6\%$, $1.2\%$, $2.4\%$, $4.8\%$, $9.6\%$, and $19.2\%$ of the highest peak
flux density of $2.79$ Jy/beam (epoch $2009.23$ as listed in Table
\ref{table:1}) for common reference. The restoring size is $1.04$\,mas $\times$
$0.45$\,mas at a position angle of $-6.4^{\circ}$. Labels mark the position of
\texttt{MODELFIT} components while unmarked components indicate non-robust
features. Time progresses from top to bottom and from left to right.}
\label{maps15}
\end{figure*}

\begin{figure}
   \centering
\includegraphics[height=0.92\textheight]{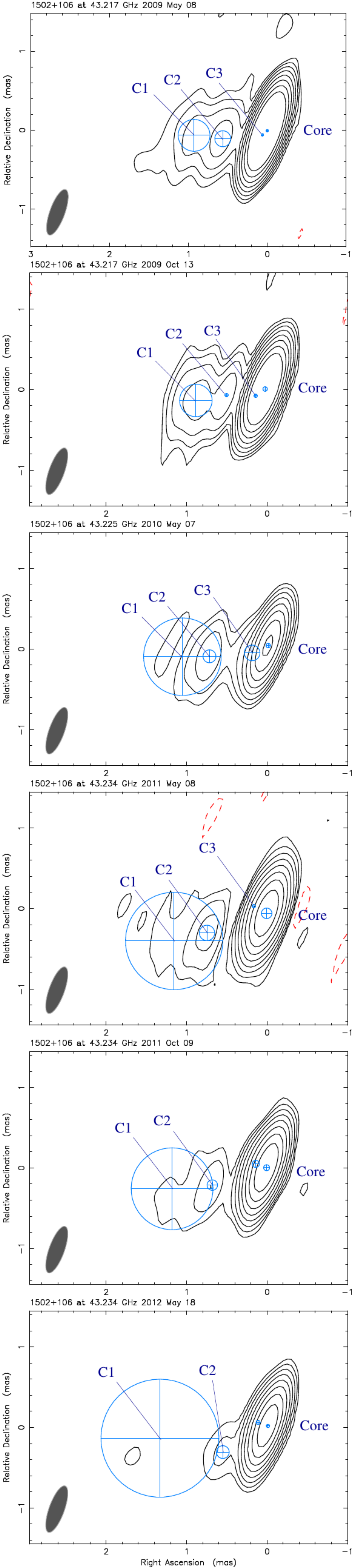}
\caption{Uniformly weighted, untapered \texttt{MODELFIT} images of PKS\,1502+106
at $43$\,GHz. Controur levels correspond to $-0.1\%$, $0.1\%$, $0.2\%$, $0.4\%$,
$0.8\%$, $1.6\%$, $3.2\%$, $6.4\%$, and $12.8\%$ of the peak flux density of
$3.47$\,Jy/beam (epoch $2009.35$, see Table \ref{table:2}) for common
reference. The restoring beam is shown at the bottom left of each image, with
FWHM $0.61$\,mas $\times$ $0.19$\, mas at a position angle of $-19.9{^\circ}$.
Unmarked components indicate non-robustly identified features.}
\label{maps43}
\end{figure}

\begin{figure}
   \centering
\includegraphics[height=0.92\textheight]{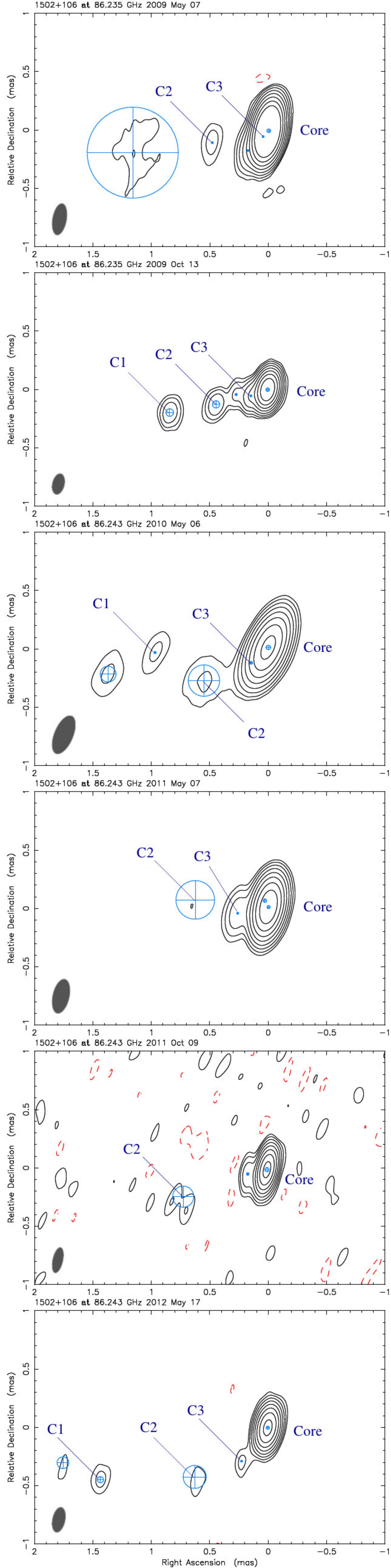}
\caption{Uniformly weighted, tapered \texttt{MODELFIT} images of
PKS\,1502+106 at $86$\,GHz. Contour levels correspond to $-0.15\%$, $0.15\%$,
$0.3\%$, $0.6\%$, $1.2\%$, $2.4\%$, $4.8\%$, $9.6\%$ and $19.2\%$ of the peak
flux density of $2.23$\,Jy/beam (epoch $2009.35$, see Table
\ref{table:3}) as a common reference. The restoring beam is shown at the bottom
left of each image. Unmarked components indicate non-robustly identified
features.}
\label{maps86}
\end{figure}

\texttt{MODELFIT} VLBI maps at 15, 43, and 86\,GHz are presented in Figs.
\ref{maps15}, \ref{maps43}, and \ref{maps86}, respectively. PKS\,1502+106
exhibits a pronounced core-dominated morphology with a continuous
one-sided jet at parsec scales.

The entire series of VLBI images -- at all frequencies -- features a distinct
region of enhanced emission responsible for the bulk of observed flux density
(see Sect. \ref{Sect:Decomp}). This brightest feature is referred to as the
core. For the purposes of our analysis we assume it stationary
for the full length of our observations and across-frequencies. This is
consistent with the fact that a putative opacity shift is smaller than our
positional accuracy (see Sect. \ref{Sect:DistanceEstimates}).

In addition to the core, the 15\,GHz jet can be decomposed into 3 to 4
distinct \texttt{MODELFIT} components at each observing epoch. At 43\,GHz the
jet is better represented by 3 components, while at 86\,GHz we have used a
number of \texttt{MODELFIT} components ranging between 2 and 4 due to varying
dynamic range of the images and intrinsic structural and flux density changes of
the source. Nevertheless, we are able to cross-identify 2 components between all
three frequencies and a third one, only between 43 and 86\,GHz. In our analysis
we use the following nomenclature: At 15 GHz, five jet features can be
identified (Fig. \ref{maps15}). The three earliest visible are labeled Ca, Cb,
and Cc. This selection is due to their estimated ejection times which are
consistent with a separation from the 15\,GHz core, prior to or close in time,
to the first observing epoch we employ (see Table \ref{15gigkin} and Sect.
\ref{kinematics} on obtaining the kinematical parameters). We use arithmetic
labeling (C1, C2, C3) for components that can be positively cross-identified
between, at least, two frequencies. More specifically, C1 and C2 are present at
15, 43, and 86\,GHz, while component C3 is only visible at the latter two
frequencies due to blending effects at 15\,GHz.

\begin{figure}
 \centering                         
 \includegraphics[width=0.6\linewidth]{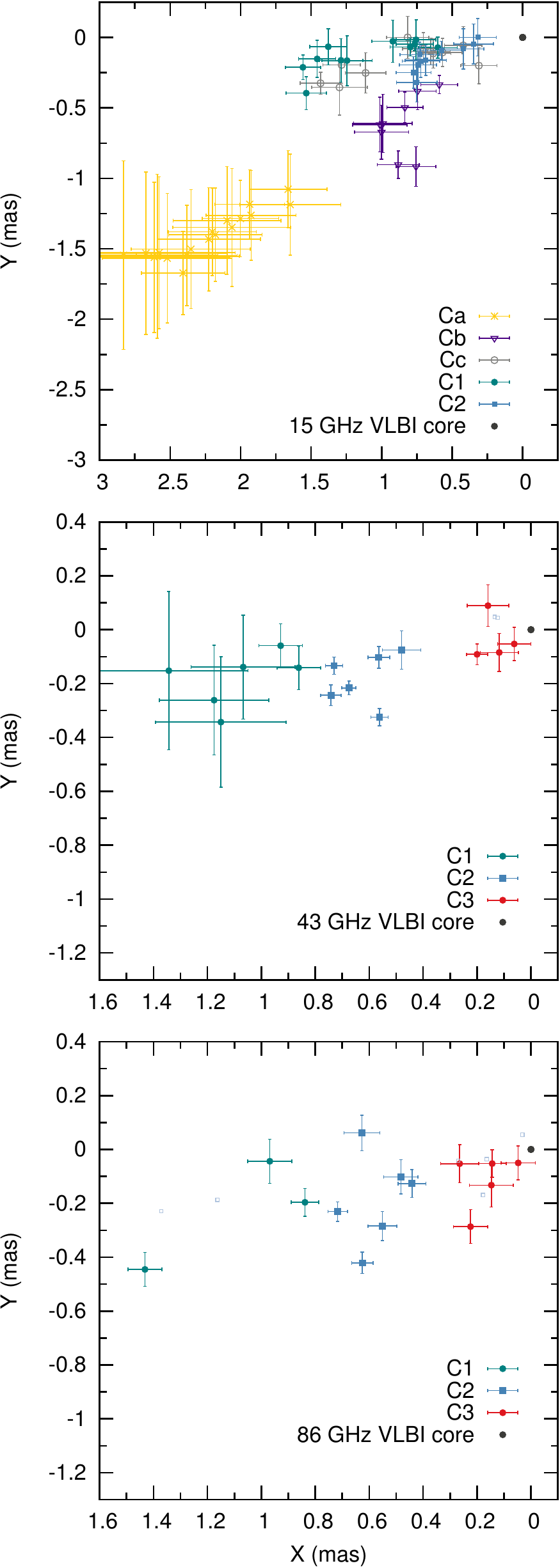}
 \caption{Sky coordinates (relative RA and DEC with respect to the
core) of all robustly identified components at $15$, $43$ and $86$\,GHz.} 
 \label{XYall}%
\end{figure}

Overall, the maximum angular extension of the jet at 15\,GHz reaches about 4
mas towards the southeast direction with the inner jet, up to
a distance of ${\sim} 2$ mas, laying at a smaller position angle (PA) and more
towards the easterly direction (see Fig. \ref{maps15}). A misalignment
between the inner and the outer jet is evident from the maps at 43 and
86\,GHz as seen in Fig. \ref{maps43} and \ref{maps86}. This is further
supported by the relative RA and DEC of fitted components shown in
Fig. \ref{XYall} at both high frequencies, where regions closer to the core
are resolved. Consequently, observations at 43 and 86\,GHz reveal the
inner jet at its highest detail.

\subsection{VLBI kinematics at 15, 43, and 86\,GHz} \label{kinematics}

\begin{figure}
 \centering                              
 \includegraphics[width=\linewidth]{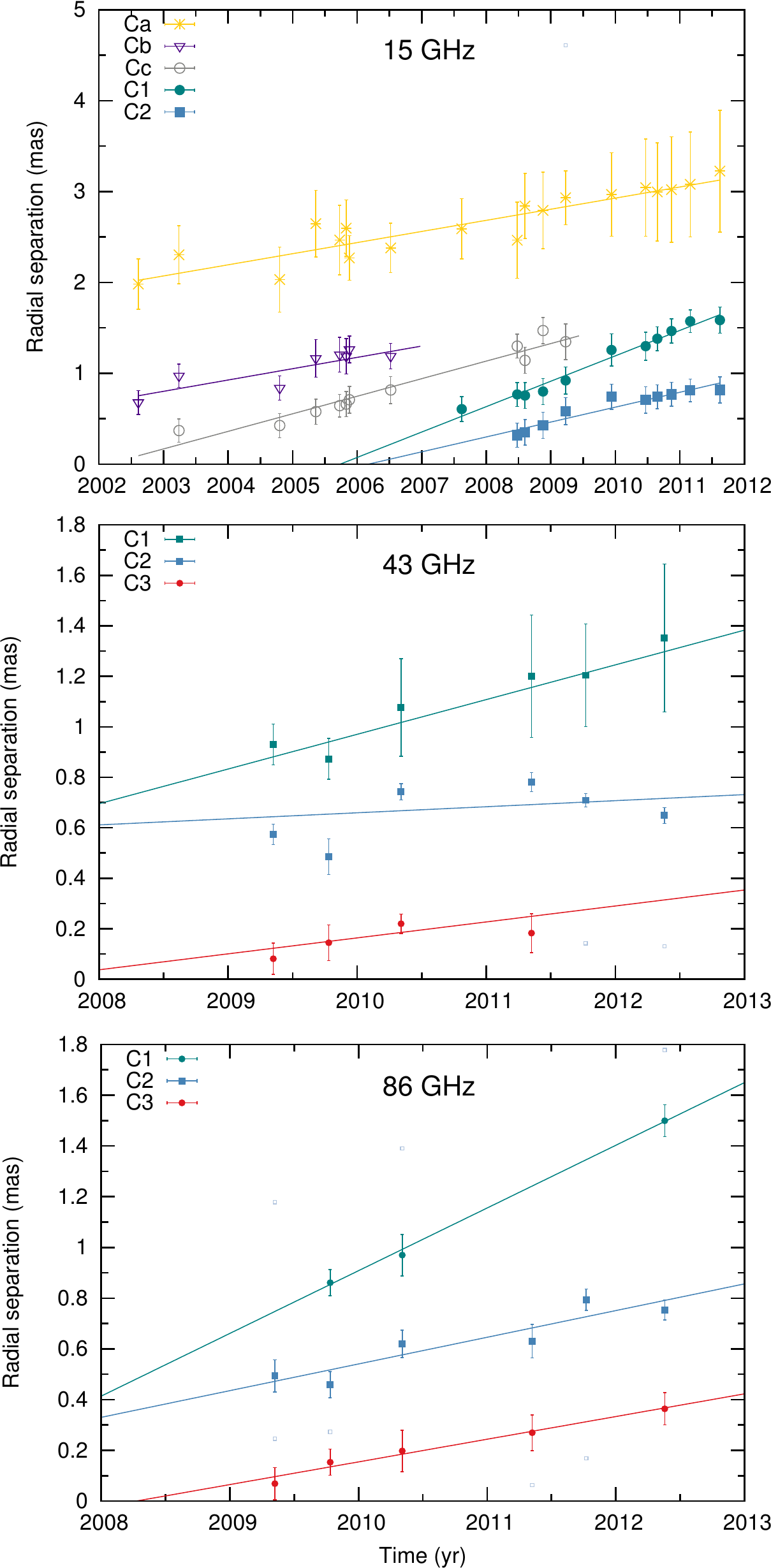}
  \caption{Temporal evolution of component radial separation from the core
at -- from top to bottom -- $15$, $43$, and $86$\,GHz. While the long-term
kinematical behavior of PKS $1502{+}106$ is visible at the lowest frequency,
$43$ and $86$\,GHz observations allow for a high resolution view towards the
inner jet of PKS $1502{+}106$. Components C1 and C2 identified across
all frequencies can be seen using the same color code.}
 \label{sepAll}%
\end{figure}

To obtain each knot's proper motion, $\mu$, through the narrow, curved
jet of PKS\,1502+106 we perform a weighted, linear least-squares fit to their
epoch-to-epoch radial separation from the respective core at each
frequency. Component radial separations from the core are shown in Fig.
\ref{sepAll} for 15, 43, and 86\,GHz respectively, where the linear fits are
also visible. The inferred values of apparent speed in mas/yr and in units of
speed of light ($\beta_{{\rm app}} = \upsilon_{\rm app}/c$), along with ejection
dates for all robust jet features are reported in Tables \ref{15gigkin},
\ref{43gigkin}, and \ref{86gigkin}.

Compared to the kinematical results of the
MOJAVE program presented in \cite{2013AJ....146..120L}, our component
identification is slightly different. It is also based in the
cross-identification with the high-frequency data but the overall picture does
not differ significantly and knot speeds along with their ejection dates are in
very good agreement. 

The apparent velocity, $\beta_{\rm{app}}$, of each component is
given by:
\begin{equation}
\beta_{{\rm app}}=\dfrac{\mu \, D_{\rm L}}{c (1+z)}
\end{equation}
with,
\[
\begin{array}{lp{0.8\linewidth}}
  \mu,       & the proper motion in rad\,s$^{-1}$, \\
  D_{\rm L}, & the luminosity distance in m, \\
  c,         & speed of light in m\,s$^{-1}$, and \\
  z,         & the redshift of the source. \\
\end{array}
\]

As evident from Fig. \ref{sepAll} and Tables \ref{15gigkin} through
\ref{86gigkin}, PKS\,1502+106 is characterized by extremely fast superluminal
motion of features within the jet. Measured apparent velocities are in the
range of ${\sim}10$ to $22\,c$ at 15\,GHz, while at 43\,GHz the range is
$2$--$11\,c$ and from $7$ to $20\,c$ for 86\,GHz. The maximum
$\beta_{\rm app}$ per frequency is 
$\beta_{\rm app}^{\rm 15\,GHz} = (22.1 \pm 1.1)$,~
$\beta_{\rm app}^{\rm 43\,GHz} = (11 \pm 3)$,~
$\beta_{\rm app}^{\rm 86\,GHz} = (19.5 \pm 0.4)$, all characterizing the
motion of superluminal feature C1.

The independently deduced kinematical models per frequency are generally in
agreement, with only a few, worthy of discussion
discrepancies. These are mainly encountered at our mid-frequency of 43\,GHz for
components C1 and C2. Here, C1 appears to be traveling downstream the jet with
only about half the speed observed at 15 and 86\,GHz, where the measurements
agree within the observational uncertainties. This is attributed to the
large positional uncertainties but also to the small number of high frequency
VLBI observations (especially at 43\,GHz) which are also limited in
$uv$-coverage. Consequently, the absence of even a single data point can
drastically affect the quality of the fit and the deduced kinematical
parameters.

\begin{figure}
 \centering
 \includegraphics[width=\linewidth]{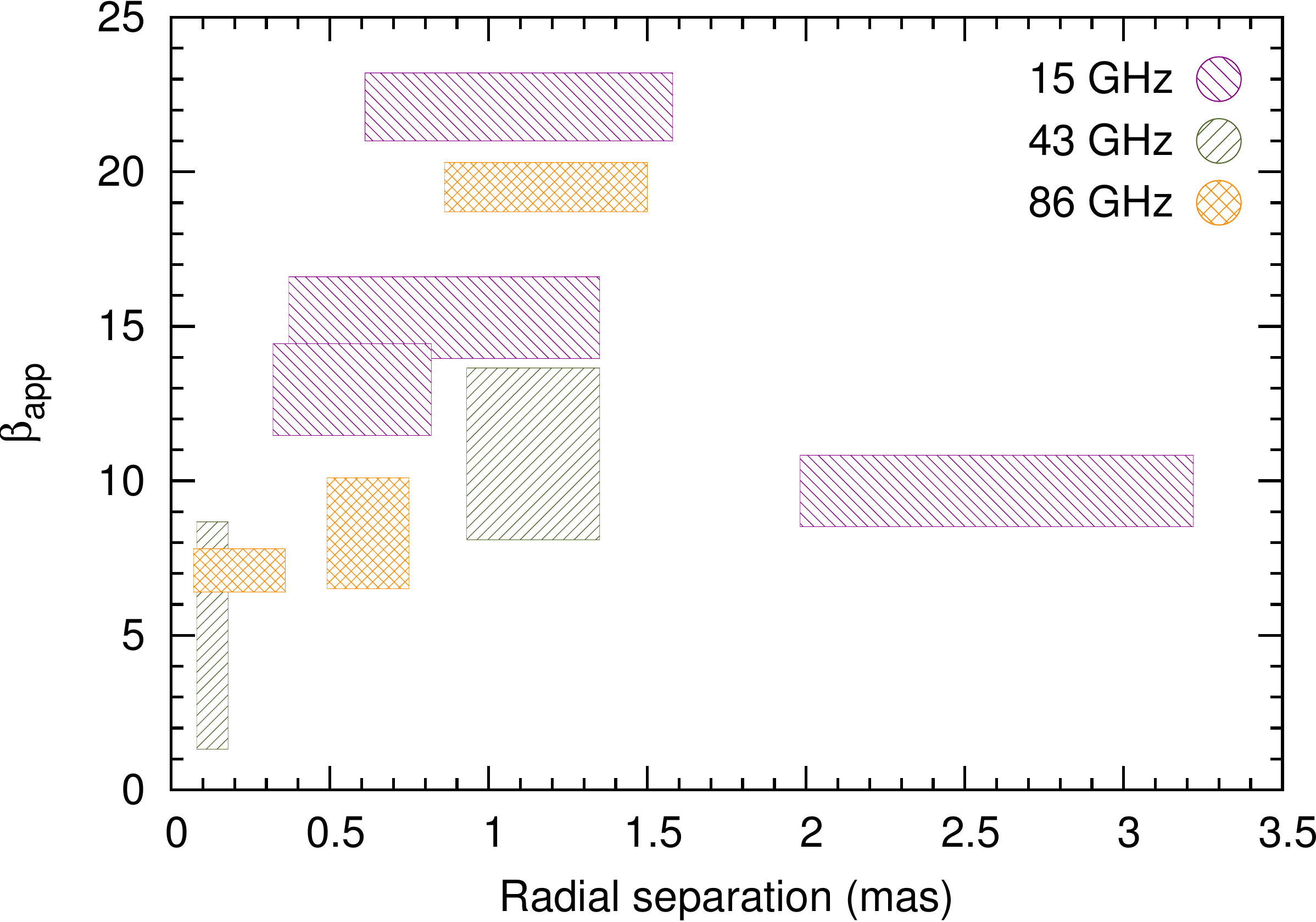}
  \caption{Apparent speed profile of superluminal components as a function of
radial separation from the core. The vertical extent of rectangles
represents the uncertainty in $\beta_{\rm app}$. Their horizontal extent
denotes the distance that each component has traversed within the jet, during
the monitoring period. A rising trend is seen up to a projected separation of
${\sim}$1.6\, mas, after which a lower apparent speed is seen.}
 \label{beta_appa}%
\end{figure}

Inferred apparent component speeds allow for a study of the $\beta_{\rm app}$
profile along the radial direction of the jet. This is presented
in Fig. \ref{beta_appa}. Up to a distance of about $1.5$\,mas from the
core the apparent speeds observed, follow an almost monotonically increasing
trend, peaking at about 1\,mas at all frequencies. Beyond 1.5\,mas
downstream, the trend seems to break with only one, though robust,
measurement at 15\,GHz for the historical component Ca having only a $\beta_{\rm
app} = (9.7 \pm 1.2)$ at a radial separation of ${\sim}2.8$\,mas, indicating a
change of physical conditions within the flow -- e.g. deceleration and/or
change of the viewing angle. In the immediate vicinity of the core, knot C3 
moves outwards with an apparent superluminal speed of $(5.0 \pm 3.7)\,c$ and
$(7.1 \pm 0.7)\,c$, at 43 and 86\,GHz respectively.

The XY (RA--DEC) position and the radial separation plots presented in Fig.
\ref{XYall} and \ref{sepAll} reveal the erratic path of
superluminal feature C2 both at 43 and 86\,GHz. The emerging pattern is
not inconsistent with, and could in fact hint towards, the scenario that C2 is
following a helical trajectory. A scenario though, that cannot be
corroborated due to the limited number of observations and the consequently low
sampling rate of the underlying motion.

Other physical effects may also contribute to small, observed differences. For
example, the small angles under which blazars are viewed, having the potential
of greatly magnifying relativistic aberration effects and the position of the
core itself which can constitute a non-stationary feature, subject to erratic
motion. However, such an effect in not clearly seen by the motion of other
superluminal components. An additional effect, as resolution increases, is that
more compact regions are picked up by the interferometer, thus shifting
the centroid of a fitted component within a larger emission region (as seen
at lower frequencies).

For those jet components that can be identified across observing
frequencies, ejection dates are again in good agreement, with the exception of
C2 whose wobbly motion does not allow a good estimation of the ejection date
at 43\,GHz. Components C1, C2 and C3 have been separated from the core in the
time period covered by the VLBI monitoring. C1 and C2 were ejected between
2005 and 2006, while C3 in ${\sim} 2008$, at a time close to the onset of the
multi-frequency outburst of PKS\,1502+106.

\begin{table}
 \centering
 \caption[]{Derived kinematical parameters of jet components at $15$\,GHz}
 \label{15gigkin}
 \begin{tabular}{l c c c }
 \hline
 \hline
 Knot    &      $\mu$         & $\beta_{\rm app}$ &        $t_{\rm ej}$  \\
 	 & $({\rm mas\,yr^{-1}})$  & ($c$) & (yr) \\
  \hline\noalign{\medskip}
     Ca  & $0.122 \pm  0.015$ & $~~9.7 \pm 1.2$  & $1986.1 \pm 2.6 $  \\
     Cb	 & $0.125 \pm  0.033$ & $~~9.8 \pm 2.6$  & $1996.6 \pm 2.2 $  \\
     Cc	 & $0.193 \pm  0.017$ & $15.3 \pm 1.3$   & $2002.1 \pm 0.4 $  \\
     C1	 & $0.280 \pm  0.014$ & $22.1 \pm 1.1$   & $2005.7 \pm 0.2 $  \\
     C2	 & $0.164 \pm  0.019$ & $13.0 \pm 1.5$   & $2006.2 \pm 0.5 $  \\
 \noalign{\smallskip}
 \hline
 \end{tabular}
\end{table}

\begin{table}
 \centering
 \caption[]{Derived kinematical parameters of jet components at 43\,GHz}
 \label{43gigkin}
 \begin{tabular}{l c c c}
 \hline\hline
 Knot    &      $\mu$	      & $\beta_{\rm app}$ &   $t_{\rm ej}$	  \\
 	 & $({\rm mas\,yr^{-1}})$  & ($c$) & (yr) \\
  \hline\noalign{\medskip}
     C1  & $0.138 \pm  0.035$ & $10.9 \pm 2.8$  & $2003.0  \pm ~~2.0$ \\
     C2	 & $0.024 \pm  0.035$ & $~~2.0 \pm 2.7$ & $1982.5  \pm 41.5$  \\
     C3	 & $0.063 \pm  0.047$ & $~~5.0 \pm 3.7$ & $2007.4  \pm ~~2.2$ \\
  \noalign{\smallskip} 
  \hline 
 \end{tabular}
\end{table}

\begin{table}
 \centering
 \caption[]{Derived kinematical parameters of jet components at 86\,GHz}
 \label{86gigkin}
 \begin{tabular}{l c c c}
 \hline\hline
 Knot    &      $\mu$	      & $\beta_{\rm app}$ &   $t_{\rm ej}$	  \\
 	 & $({\rm mas\,yr^{-1}})$  & ($c$) & (yr) \\
  \hline\noalign{\medskip}
     C1	 & $0.247 \pm  0.010$ & $19.5 \pm 0.8$  & $2006.3  \pm 0.2$   \\
     C2	 & $0.105 \pm  0.023$ & $~~8.3 \pm 1.8$ & $2004.9  \pm 1.4$   \\
     C3	 & $0.090 \pm  0.009$ & $~~7.1 \pm 0.7$ & $2008.3  \pm 0.3$   \\
  \noalign{\smallskip}
  \hline
 \end{tabular}
\end{table}

\subsection{Radio flux density decomposition}\label{Sect:Decomp}

\begin{figure*}
 \centering
 \includegraphics[width=0.9\textwidth]{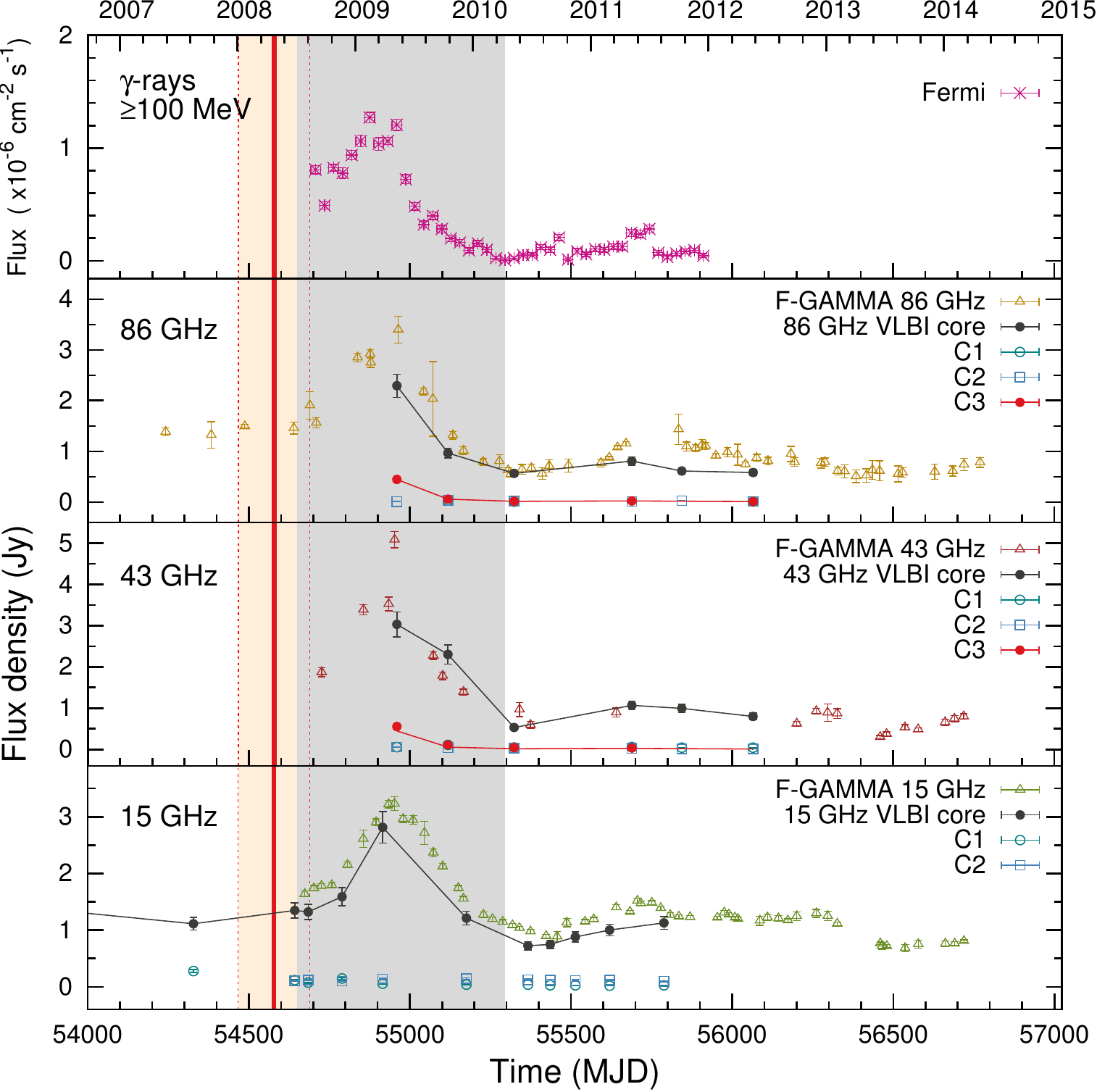}
  \caption{Light curves of PKS $1502{+}106$. From top to bottom are
shown: the monthly-binned \textit{Fermi}/LAT $\gamma$-ray light curve at
energies $E > 100$\,MeV; F-GAMMA single-dish radio light curve at $86$\,GHz
with the core and component light curves from the VLBI flux density
decomposition; same as before but at $43$\,GHz and (lower panel) at $15$\,GHz.
Note the VLBI knot C3, only visible at the two highest frequencies, at its
decaying flux density phase. The gray shaded area delineates the full duration
of the flare ${\sim} 650$ days. The red shaded area marks the estimated ejection
interval of knot C3 with the red solid line designating its $t_{\rm ej}$.}
 \label{LCs_all}%
\end{figure*}

The flare in PKS\,1502+106 is dominated by a single, clearly visible
event from $\gamma$-ray energies down to radio frequencies, between
MJD\,$54689$ and MJD\,$55295$. The upper panel of Fig. \ref{LCs_all} features
the \textit{Fermi}/LAT $\gamma$-ray light curve and following panels depict the
total intensity radio data, both from the filled-aperture F-GAMMA observations
and from cm/mm-VLBI monitoring. The total, single-dish flux density is
decomposed into individual VLBI core and jet-component light curves. We show
those of the positively cross-identified components C1, C2, and C3, omitting the
historical components, at 15\,GHz, and any non-robust jet features.

The core dominates the total flux density at VLBI scales at all frequencies.
Radio flux density decomposed into core and \texttt{MODELFIT} components
accounts for the largest fraction of the total single-dish flux density and
follows nicely the flare evolution. Only in few cases, the sum of flux density
of all \texttt{MODELFIT} components deviates from the total single-dish flux
density. This is attributed to the low cadence of mm-VLBI monitoring and the
source's extreme variability behavior. The presence of substructure in the
well-sampled F-GAMMA and $\gamma$-ray light curves provides evidence for such
fast variability.

Our mm-VLBI monitoring reveals that the radio flare, for its whole duration, is
dominated by emission from the core, which at 15\,GHz has varied in flux
density by a factor of $\sim$3 with a time scale of approximately 2 years (see
Fig. \ref{LCs_all}). The same behavior persists also at 43 and 86\,GHz  where
its flux density varies by a factor of $\sim$6--7 and $\sim$4--5, respectively,
within our GMVA monitoring period.

At 43 and 86\,GHz, where resolution allows, it is clear that the flare is not an
attribute of the core only. Here, the most recently ejected component C3 appears
to be in a decaying phase already during the first mm-VLBI observing epoch, but
still shows significantly higher flux densities compared to all later epochs
(see middle panels of Fig. \ref{LCs_all} and Fig. \ref{compLCsall}). Despite
the limited number of data for C3, it can be clearly
seen that -- apart form the core -- it is the only moving jet feature that is
flaring, with its flux density dropping by a factor of ${\sim} 20$, at both
frequencies. For completeness, all individual component light curves are shown
in the Appendix.

\subsection{Spectra of individual knots}\label{sect:spectra}

\begin{figure*}
   \centering
   \includegraphics[width=0.95\textwidth]{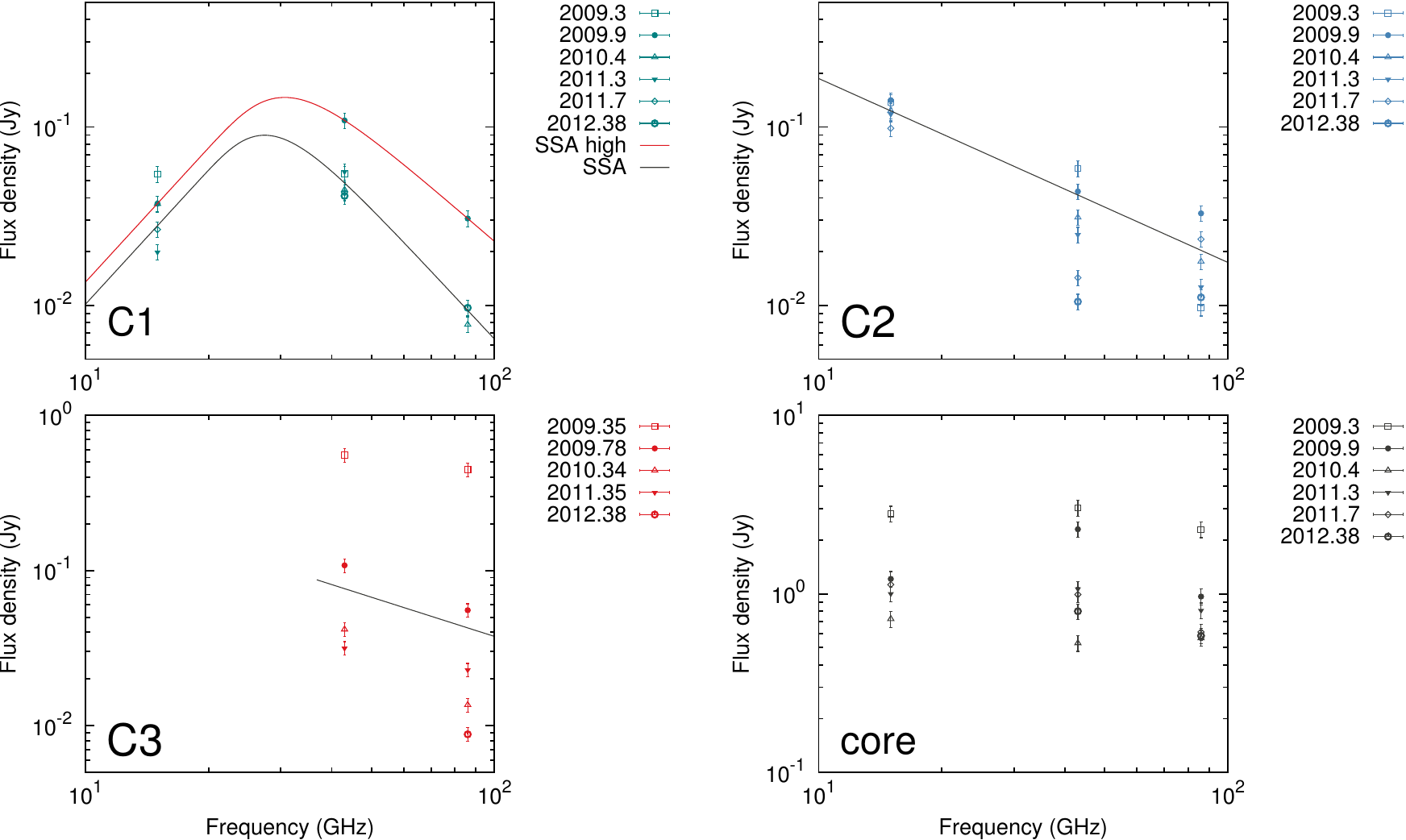}
   \caption{Three-point spectra of knots C1, C2, the VLBI core and
their temporal evolution. For C3 spectra are only available at $43$\,GHz and
$86$\,GHz since it is blended at $15$\,GHz. Black lines represent
the time-averaged spectra of components and their slopes
are reported in Table \ref{table:spInd}. For component C1 the fitted SSA model
spectra are also shown. The red curve is a SSA fit during its highest state,
while the black corresponds to the mean SSA spectrum.}
   \label{compSpectraAll}
\end{figure*}

While our high-frequency GMVA data at 43 and 86\,GHz are concurrent, this is
not the case for the 15\,GHz VLBA data. In order to assemble
quasi-simultaneous spectra of individual superluminal components within the jet
of PKS\,1502+106, we select the 15\,GHz epochs closest in time to each of the
six 43/86\,GHz GMVA observing runs. The quasi-simultaneous data forming each of
the six spectra are labeled, \textrm{I} to \textrm{VI}, in Tables
\ref{table:2} through \ref{table:1}. Assembled spectra are shown in Fig.
\ref{compSpectraAll}. The mean separation between the 15\,GHz and 43/86\,GHz
observing epochs composing them is ${\sim} 51$ days.

Components C1 and C2 are positively cross-identified between
all three observing frequencies (Sect. \ref{pcScale}) and hence
quasi-simultaneous spectra were assembled for those. Knot C1 is
the only sufficiently sampled component that shows evidence of a synchrotron
self-absorbed spectrum (SSA) featuring a spectral break at the turnover
frequency, $\nu_{\rm m}$ and flux density $S_{\rm m}$. In order to obtain
$S_{\rm m}$ and $\nu_{\rm m}$, along with the optically-thin spectral index we
fit a model SSA spectrum of the following form
\citep[e.g.][]{2000A&A...361..850T}:
\begin{equation}
 S_{\nu} = S_{\rm m} \left( \dfrac{\nu}{\nu_{\rm m}} \right)^{\alpha_{\rm t}}
 \dfrac{ 1 - \exp[-\tau_{\rm m}(\nu / \nu_{\rm m})^{\alpha - \alpha_{\rm t}}]
}{1 - \exp (-\tau_{\rm m}) }
\end{equation}
where:
\[
\begin{array}{lp{0.8\linewidth}}
 \nu_{\rm m},    & is the turnover frequency in\,GHz,    \\
 S_{\rm m},      & is the turnover flux density in Jy,   \\
 \tau_{\rm m},   & $\sim 3/2 \left( \sqrt{1-\dfrac{8 \alpha}{3\alpha_{\rm
t}}} - 1 \right)$ is the optical depth at $\nu_{\rm m}$, \\
 \alpha_{\rm t}, & is the spectral index of the optically thick part of
the emission, and                                        \\
 \alpha,         & is the optically-thin spectral index. \\
\end{array}
\]
The spectral index for the optically thick part is fixed to the canonical value
$\alpha_{\rm t} = 5/2$. We fit the observed spectra of C1 simultaneously and
obtain $S_{\rm m}= (0.08 \pm 0.01)$\,Jy, $\nu_{\rm m} = (33.2 \pm 2.0)$\,GHz and
$\alpha = -(2.5 \pm 0.4)$, as the average spectral index. The spectral index
characterizing knot C1 is arguably very steep. This could indicate that
the component at the highest frequency is resolved-out, as a result of high
resolution (i.e. small observing beam), leading to an underestimation of its
flux density. An additional uncertainty in $\alpha$ of up to 15\% is also to be
expected, arising from the different $uv$-coverages
\citep[e.g.][]{2013A&A...557A.105F}. To obtain the same parameters
during the highest state of the knot, the fit is performed once more, at epoch
2009.9 only. Resulting parameters are as follows: $S_{\rm m}=0.133$\,Jy,
$\nu_{\rm m} = 36.7$\,GHz and $\alpha = -1.98$. These were used for the
calculation of the magnetic field from SSA in Sect. \ref{Sect:Bssa}.

For components C2 and C3 showing no sign of a SSA break, a
power-law fit is performed separately at each observing epoch and the mean
optically thin spectral index $\left< \alpha \right>$, characterizing the
time-averaged spectrum, along with its standard error are reported in Table
\ref{table:spInd}.

C2 is also a well-sampled feature that during the whole length of our VLBI
monitoring is characterized by a steep optically-thin spectrum with a mean
power-law index of $\left< \alpha \right>  = -(1.0 \pm 0.3)$ and a
peak at ${\le} 15$\,GHz. However, the situation is less constrained for
component C3 that is only seen at $43$ and $86$\,GHz due to the resolution
limit of the $15$\,GHz observations. Here, the analysis yielded an
optically-thin, two-point spectral index $\left< \alpha \right>  = -(0.8 \pm
0.3)$.

The core is the most compact, stationary feature at each frequency and it is
characterized by a flat spectrum; i.e. $\alpha \ge -0.5$. However, at epoch
$2009.9$ it appears to exhibit a SSA turnover at $\nu_{\rm m} \sim 43$\,GHz.
The mean spectral index of the core is found to be $\left< \alpha \right>  =
-(0.22 \pm 0.07)$, indicative of unresolved substructure.

\begin{table}
\caption{Spectral indices for components C1, C2, C3, and the core as obtained
by SSA fitting for C1 and a single power-law fit for C2 and C3.} 
\label{table:spInd}
\centering          
\begin{tabular}{l c c}
\hline
\hline
Knot      & $\left< \alpha \right> $ \\
\hline\noalign{\medskip}
C1        &    $-2.5 \pm 0.4$   \\
C2        &    $-1.0 \pm 0.3$   \\
C3        &    $-0.8 \pm 0.3$   \\
Core      &    $-0.22 \pm 0.07$   \\
\noalign{\smallskip}
\hline                  
\end{tabular}
\end{table}

%
\section{Inferred physical parameters}

In the present section, we deduce physical parameters of the jet of
PKS\,1502+106, given the phenomenological characteristics presented in Sect.
\ref{Sect:Phenomenology}. We discuss the Doppler factor, viewing angle, and
brightness temperature distribution along the jet, and estimate magnetic field
strengths.

\subsection{Doppler and Lorentz factor estimates}\label{Sect:Dfctr}

As shown in Sect. \ref{kinematics}, the highest apparent speed measured in the
jet of PKS\,1502+106 is 
$\beta_{\rm app} \sim 22$ and corresponds to that of component C1 at
15\,GHz. From here we can readily estimate the minimum Lorentz
factor that characterizes the flow \citep[see for example the review
by][]{1995PASP..107..803U}. It follows that $\gamma \geq \gamma_{\rm min}$ for
the jet with the minimum Lorentz factor given by
\begin{equation}
    \gamma_{\rm min} = \sqrt{\beta_{\rm app}^{2} + 1},
\end{equation}
and for the observed maximum apparent velocity we obtain $\gamma_{\rm min} =
22.1$. Consequently, if the jet is viewed at the critical angle (see Sect.
\ref{Sect:theta}), Doppler factors can be estimated too, since
$\delta_{\beta_{\rm app},\,C1} \sim \beta_{\rm app,\,C1} = 22.1$ at 15
GHz. For knots C2 and C3, it is $\delta_{\beta_{\rm app}} \sim \gamma_{\rm
min,\,C2} = 9.3$ and $\delta_{\beta_{\rm app}} \sim \gamma_{\rm min,\,C3} =
7.2$, from the 86\,GHz observations.

In addition to the Doppler factor estimate obtained from the minimum Lorentz
factor, under the valid assumption of causality one can arrive to another
estimate of the Doppler factor, $\delta_{\rm var}$, using the temporal variation
of the flux density of individual components. For details on this method and for
a full derivation see \cite{2005AJ....130.1418J} and
\cite{2013Ap&SS.348..193O}, respectively. The variability Doppler
factor is given by
\begin{equation}
\delta_{\rm var}=\dfrac{d_{{\rm eff}} \, D_{\rm L}}{c \Delta t_{{\rm var}}
\,(1+z)},
\end{equation}
where:
\[
\begin{array}{lp{0.8\linewidth}}
 d_{{\rm eff}},  & is the effective angular size of the component in rad and \\
 \Delta t_{\rm var}, & the variability time scale, $[{=}dt/\ln (S_{\rm
max}/S_{\rm min})]$, with $dt$ the time between $S_{\rm max}$ and $S_{\rm
min}$\\
\end{array}
\]

Components C1 and C3 can be used for this purpose. Both components are
unresolved; that is, being smaller than the beam size at the epoch of their
highest respective flux densities at all three observing frequencies. In all
cases (for C1 at 15\,GHz and for C3 at 43 and 86\,GHz) we have used the
minor axis of the restoring beam as the upper limit for the component angular
size. This selection is based on the fact that the minor axis is positioned
almost parallel to the jet axis, thus allowing the highest resolution to be
achieved along it. Specifically, for the higher frequencies and component C3 the
minor axis of the 86\,GHz beam was selected, since it yields the most stringent
upper limit to its size.

A simple inspection of the light curves indicates that component
flux density variations are not adequately sampled, while the point of maximum
flux density most probably precedes our first observing epoch.
Nevertheless, based on the comparison with the single-dish data (Sect.
\ref{Sect:Decomp}), this point cannot be too displaced (in time), nor much
higher (in flux density) from the first data point in the mm-VLBI light curves.
We therefore employ the method, since useful nominal values can be obtained. We
calculate $\Delta t_{\rm var}$, with $dt$ and $S_{\rm max}$ obtained directly
from the light curves of individual components.

Along the same argumentation an estimate of the Lorentz factor can be obtained
by
\begin{equation}
\gamma_{\rm var} = \dfrac{ \beta_{\rm app}^{2} + \delta_{\rm var}^{2} + 1 }{ 2
\delta_{\rm var} }.
\end{equation}

Results of the calculations are summarized in Table \ref{table:4}.
Component C1 is characterized by the highest variability Doppler
($\delta_{\rm var} \sim 51.4$) and Lorentz ($\gamma_{\rm var} \sim 30.5$)
factors. All figures are consistent with the estimates of the minimum Lorentz
factor, $\gamma_{\rm min}$, presented previously. Knot C3 is also found to have
consistent behavior at both high frequencies with $\delta_{\rm var}^{\rm
43\,GHz} \sim 12.1$ and $\gamma_{\rm var}^{\rm 43\,GHz} \sim 7.1$, while at
$86$\,GHz the estimation yields $\delta_{\rm var}^{\rm 86\,GHz} \sim 15$ and
$\gamma_{\rm var}^{\rm 86\,GHz} \sim 9.2$.

An additional Doppler factor estimate can be obtained from the relative
strength of the synchrotron self-absorption and equipartition magnetic fields.
This is discussed in Sect.
\ref{Sect:Bssa}.

\begin{table}
\small
\caption{Physical parameters estimated using variability arguments.}  
\label{table:4}      
\centering          
\begin{tabular}{l c c c c c c c}
\hline
\hline
Knot  & $\nu$  & $dt$ & $d_{\rm eff}$ & $\Delta t_{\rm var}$ & $\delta_{\rm var}$ &	 $\gamma_{\rm var}$	 & $\theta_{\rm var}$ \\
      & (GHz)  & (yr) &      (mas)    &    (yr) 	&		     &  			 &   ($^{\circ}$)     \\
\hline\noalign{\medskip}
C1    & 15     & 3.6  & ${\leq} 0.9$  &    $1.3$  &   $51.4$  &  $30.5$ & $0.8$ \\ 
C3    & 43     & 1.9  & ${\leq} 0.1$  &    $0.7$ &    $12.1 $ &  $7.1$ &  $3.4$ \\ 
C3    & 86     & 1.9  & ${\leq} 0.1$  &    $0.7$ &    $15.0 $ &  $9.2$ &  $2.9$ \\
\noalign{\smallskip}
\hline                  
\end{tabular}
\tablefoot{Columns from left to right:
(1) knot designation;
(2) observing frequency;
(3) elapsed time between $S_{\rm max}$ and $S_{\rm min}$ (C1:
2007.6--2011.2 and C3: 2009.4--2011.3);
(4) effective size;
(5) variability time scale;
(6) Doppler factor;
(7) Lorentz factor and
(8) viewing angle.}
\end{table}

\subsection{The viewing angle towards PKS\,1502+106}\label{Sect:theta}

Using the observed apparent speed we can set constraints to the aspect angle
under which a jet is viewed in two ways.

First, by calculating the critical angle which results in the observed
$\beta_{\rm app}$ with the minimum $\gamma$. The apparent speed is given by
\citep{1966Natur.211..468R}
\begin{equation}
 \beta_{\rm app} = \dfrac{\beta \sin \theta}{1 - (\beta \cos \theta)},
 \label{apparent}
\end{equation}
where $\beta$ is the intrinsic velocity in units of $c$ and $\theta$
is the viewing angle. Using the definition of the Doppler factor, 
$\delta = [\gamma (1- \beta \cos \theta) ]^{-1}$, Eq. \ref{apparent} can be
rewritten as
\begin{equation}
 \beta_{\rm app} = \beta \gamma \delta \sin \theta,
 \label{apparent2}
\end{equation}
and $\beta_{\rm app}$ is maximized (i.e. $\beta_{\rm app,\,max} = \beta \gamma$)
for $\sin \theta_{\rm c} = \gamma^{-1}$ or
$\sin \theta_{\rm c} = \delta^{-1}$, since for this angle $\theta_{\rm c}$, 
$\gamma = \delta$.
With the minimum Lorentz factor, $\gamma_{\rm min} = 22.1$, calculated above
we arrive to the following figure for the critical angle:
\begin{equation}
    \theta_{\rm c} = \arcsin \left( \dfrac{1}{\gamma_{\rm min}} \right) =
2.6^{\circ}.
\end{equation}

We can also estimate the viewing angle towards the source, based on
causality arguments. This ``variability viewing angle'', $\theta_{\rm var}$, is
given by
\begin{equation}
\theta_{\rm var} = \arctan \left(\dfrac{ 2 \beta_{\rm app} }{ \beta_{\rm
app}^{2} + \delta_{\rm var}^{2} - 1 } \right).
\end{equation}
This analysis was performed for two components with significant variability
(see Fig. \ref{compLCsall}) and yields viewing angles for each component
separately. The results are shown in the last column of Table \ref{table:4}. For
the fastest superluminal knot, C1, we obtain the smallest 
viewing angle $\theta_{\rm var} \sim 0.8^{\circ}$. For C3 we obtain consistent
values at $43/86$\,GHz of ${\sim} 3.4^{\circ}$ and ${\sim} 2.9^{\circ}$,
respectively.

Interestingly, components C1 and C3 are traveling with very different apparent
speeds (Sect. \ref{kinematics}) at two different regions of the jet -- i.e. in
the outer ($r > 1$\,mas) and inner jet ($r < 0.5$\,mas), respectively. This
fact, in combination with the difference of their inferred Doppler factors
hints towards a ``two-region scenario'', wherein physical conditions are
intrinsically different. This possibility is further discussed in the following.

\subsection{Opening angle of the jet}\label{Sect:opAng}

In the present section we investigate the apparent and intrinsic opening angles
of the jet, $\phi_{\rm app}$ and $\phi_{\rm int}$, respectively employing two
different approaches. First, we use all \texttt{MODELFIT} components at all
three frequencies, and perform a simultaneous linear fit to their sizes in order
to obtain the jet opening rate. In the second approach we deduce $\phi_{\rm
app}$ from each individual \texttt{MODELFIT} component based on its size and
kinematical characteristics.

\subsubsection{The $\phi_{\rm int}$ from simultaneously fitting the effective
sizes}

\begin{figure}
 \centering
 \includegraphics[width=\linewidth]{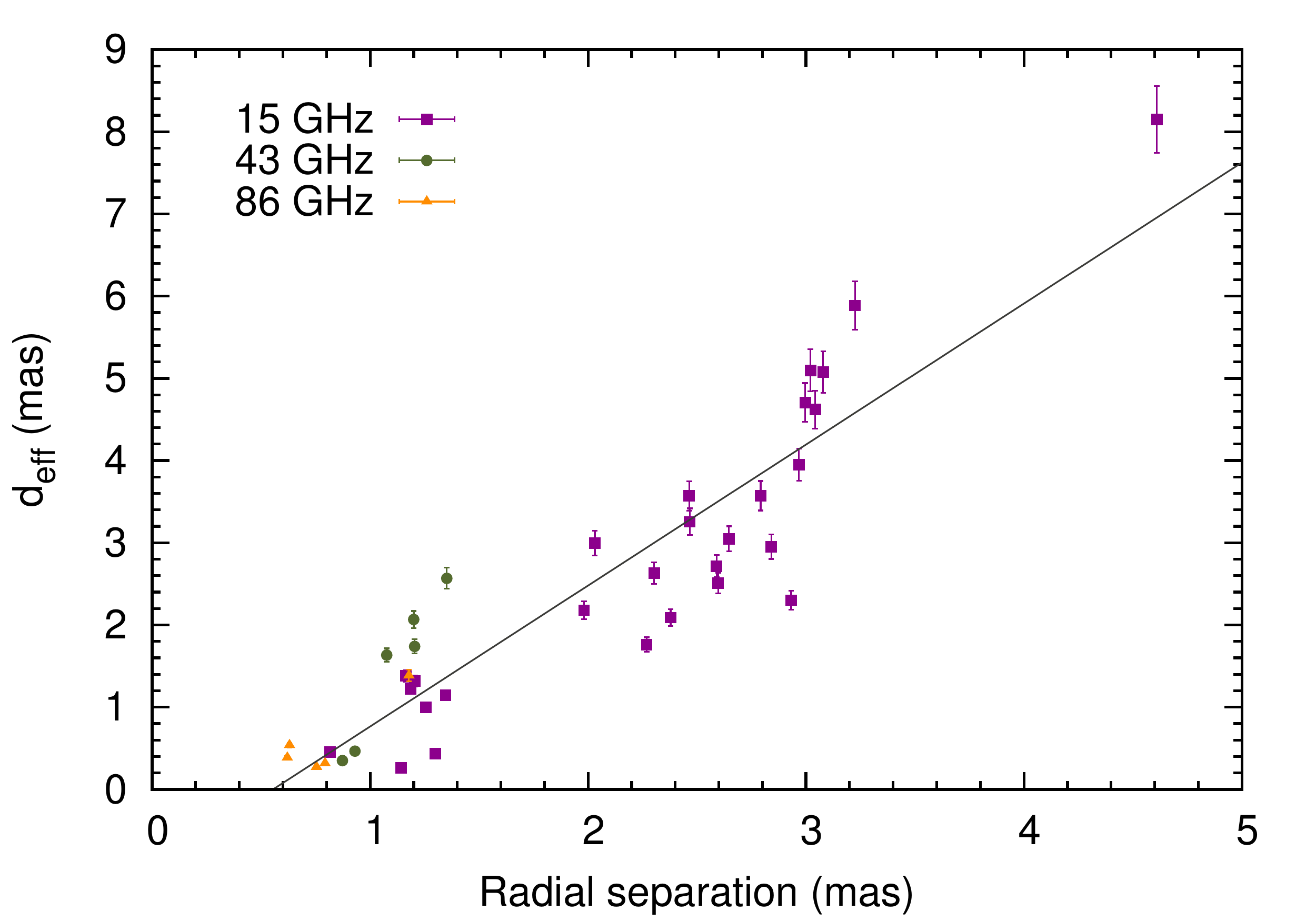}
  \caption{Effective deconvolved size of all resolved jet components with
respect to radial separation from the core, at all frequencies. A line
represents the best fit to the data.}
 \label{OPang}%
\end{figure}

In Fig. \ref{OPang} the effective deconvolved size, $d_{{\rm eff}} =
1.8\,\mbox{FWHM}_{\rm d}$ with $\mbox{FWHM}_{\rm d} = \left(
\mbox{FWHM}^{2}-b_{\phi}^{2} \right)^{1/2}$, of all resolved \texttt{MODELFIT}
components at all three frequencies is shown as a function of radial separation
from the core. The factor 1.8 accounts for the fact that a Gaussian with FWHM
equal to $d$, roughly corresponds to a circle of angular diameter $1.8\,d$.
We fit $d_{{\rm eff}}$ with respect to radial separation from the core, using a
simple linear model assuming a constant jet opening angle. The least-squares fit
yields a slope of $\left( 1.7 \pm 0.1 \right)$ translating to an apparent
half-opening angle of $(40.4 \pm 1.7)^{\circ}$ and an apparent full-opening
angle of $\phi_{{\rm app}} = (80.8 \pm 3.4)^{\circ}$. The de-projected opening
angle is then given by
\begin{equation}
    \phi_{\rm int} = \phi_{\rm app} \, \sin \theta,
    \label{intOpAng}
\end{equation}
which for $\theta = \theta_{\rm c} = 2.6^{\circ}$ (see Sect.
\ref{Sect:theta}) yields $\phi_{\rm int} = (3.7 \pm
0.2)^{\circ}$.

\subsubsection{The $\phi_{\rm int}$ from individual knots}

We performed the analysis on the $uv$-plane, utilizing the component size, ${\rm
FWHM}_{\rm  d}$, and radial distance, $r$, from the core. First, we calculate
the opening angle corresponding to each \texttt{MODELFIT} component separately.
Assuming the component filling the entire jet cross section, $\phi_{\rm app}$ is
given by
\begin{equation}
    \phi_{\rm app}^{\rm comp} = 2  \arctan  \left(
\dfrac{1.8\,{\rm FWHM}_{\rm  d}}{2 r} \right).
\end{equation}
Then, we average over effective size and distance, thus obtaining the
average opening angle of a given component at a mean radial separation from the
core. The intrinsic opening angle of the jet, making use of the
critical viewing angle for each component, is given by Eq. \ref{intOpAng}. The
results are shown in Table \ref{table:5}. The mean intrinsic opening angle for
the jet, deduced by averaging the $\left< \phi_{\rm int} \right> $ of all
components is $(3.9 \pm 1.0)^{\circ}$.

\begin{table}
\caption{Apparent and intrinsic jet opening angles deduced from individual
\texttt{MODELFIT} components.} 
\label{table:5}      
\centering          
\begin{tabular}{l c c c c c c}
\hline
\hline
Knot & $\nu$  & Epochs  &  $\left< r \right>$ & $\theta_{\rm c}$ & $\left<\phi_{\rm app} \right>$ & $\left< \phi_{\rm int} \right>$ \\
     & (GHz)  &          &         (mas)       &   ($^{\circ}$)   & ($^{\circ}$)
 	       &          ($^{\circ}$)	    	 \\
\hline\noalign{\medskip}
Ca & 15  & 19	& 2.6  & 5.9 & 63.7  & 6.5 \\      
Cb & 15	 & 3	& 1.2  & 5.8 & 57.7  & 5.8 \\
Cc & 15	 & 3	& 1.1  & 3.8 & 30.2  & 2.0 \\
C1 & 15	 & 2	& 1.3  & 2.6 & 31.3  & 1.4 \\
\noalign{\medskip}
C1 & 43  & 6  & 1.1   & 6.2 & 60.9  & 6.6  \\
\noalign{\medskip}
C2 & 86   & 4   & 0.7 & 2.9 & 31.0  & 1.6  \\
\noalign{\smallskip}
\hline                  
\end{tabular}
\tablefoot{Columns from left to right:
(1) knot designation;
(2) number of epochs used for the calculation where the component is resolved;
(3) mean separation from the core;
(4) critical viewing angle according to component's speed;
(5) mean apparent opening angle and
(6) intrinsic opening angle.}
\end{table}

\subsection{Radial distribution of the brightness temperature}\label{Sect:Tb}

The apparent brightness temperature of a given VLBI component in the source rest
frame is given by:
\begin{equation}
T_{\rm b} = 1.22 \times 10^{12} \, \dfrac{S_{\nu}}{\nu^{2} \, d_{{\rm eff}}^{2}}
\, (1+z) ~~~~(K),
\end{equation}
where:
\[
\begin{array}{lp{0.8\linewidth}}
 S_{\nu},       & is the component's flux density in Jy, \\
 d_{{\rm eff}}, & the effective size of the emitting region in mas, and \\
 \nu,           & the observing frequency in GHz. \\
\end{array}
\]

\begin{figure}
 \centering
 \includegraphics[width=\linewidth]{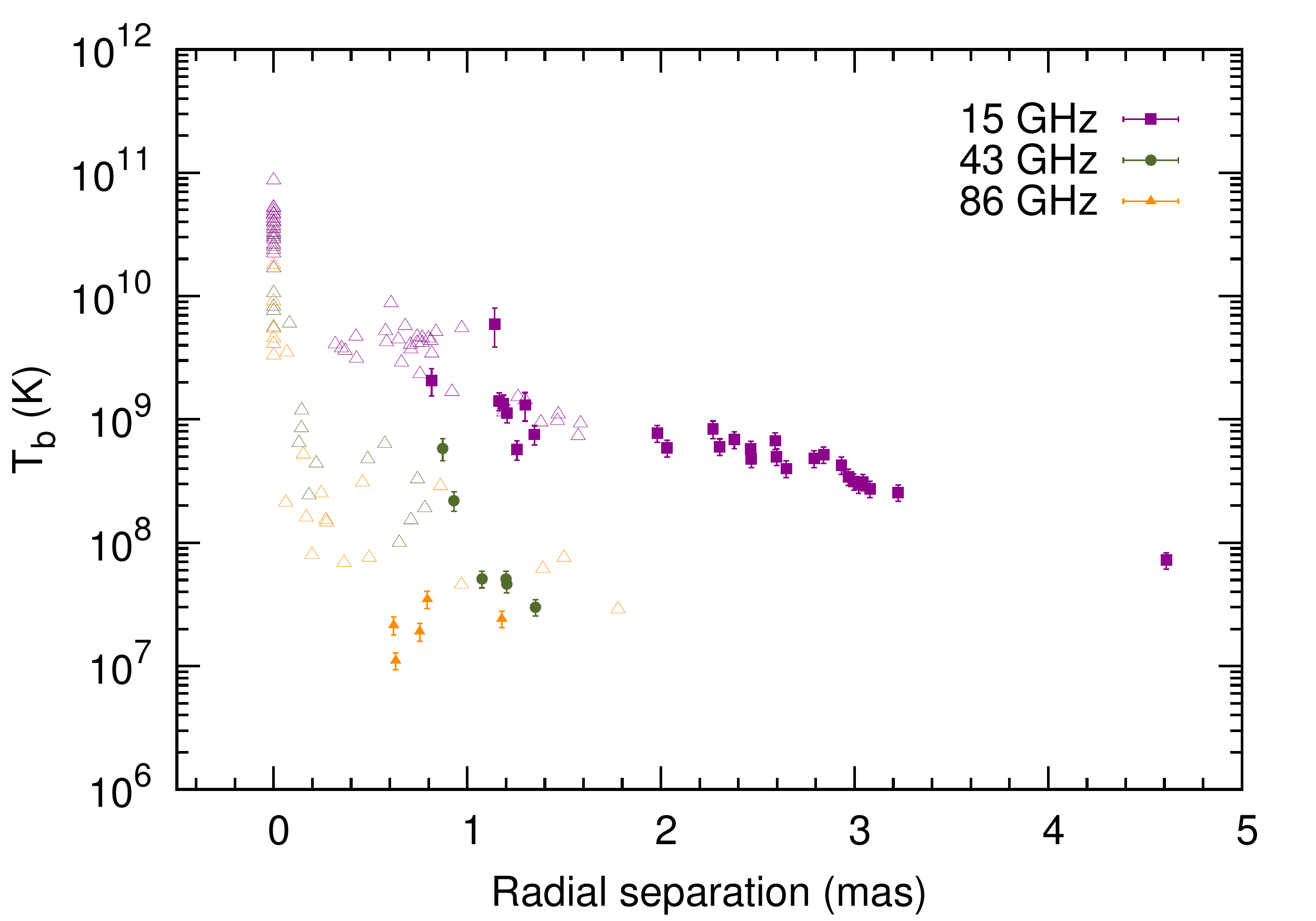}
  \caption{Brightness temperature, $T_{\rm b}$, versus the radial separation
from the core at all observing frequencies indicated by the color code. Filled
symbols represent $T_{\rm b}$ measurements, while open
triangles represent lower limits at all frequencies.}
 \label{Tb2}%
\end{figure}

In Fig. \ref{Tb2}, we present the radial distribution of brightness
temperatures along the jet of PKS\,1502+106. The $T_{\rm b}$ is
frequency-dependent, with higher values
at the lowest frequencies. Observed brightness temperatures approach and mostly
exceed the equipartition limit \citep{1994ApJ...426...51R} of about $5 \times
10^{10}$\,K for the core, at all frequencies (see Fig. \ref{Tb2}). The core is
at all times unresolved and thus the $T_{\rm b}$ values represent lower limits.
The same trend of high $T_{\rm b}$ persists throughout the first mas
downstream of the core, with a few components being resolved. Specifically, in
the regions
\begin{itemize}
 \item between $0$ and  $1$\,mas, high $T_{\rm b}$ is seen at all
frequencies, evidenced from lower limits, with a handful of measured values of
lower $T_{\rm b}$, as observing frequency increases. Those lie in the range
from $10^{7}$\,K at 86\,GHz, to slightly above $10^{9}$\,K at 15\,GHz.

\item Between $1$ and $2$\,mas, the division persists, indicative of the
sampling of different portions of the jet. Here the measurements at 15\,GHz are
clustered in the range $10^{9}$--$10^{10}$\,K at 15\,GHz, while at both high
frequencies the $T_{\rm b}$ is significantly lower, $T_{\rm b} \ge 10^{7}$\,K.

\item Beyond $2$\,mas, the jet is observable only at 15\,GHz
and the  distribution of $T_{\rm b}$ is characterized by a decreasing trend from
$10^{9}$\,K to $10^{8}$\,K at a distance of $4.6$\,mas.
\end{itemize}

The mean -- over time -- core brightness temperature at our three observing
frequencies is:
$\left<T_{\rm b,\,core}^{\rm 15\,GHz}\right> \ge (3.8\pm1.5) \times 10^{10}$\,K,
$\left<T_{\rm b,\,core}^{\rm 43\,GHz}\right> \ge (1.5\pm1.0) \times 10^{10}$\,K,
 and
$\left<T_{\rm b,\,core}^{\rm 86\,GHz}\right> \ge (7.4\pm5.0) \times 10^{9}$\,K.

\subsection{Magnetic field estimates}\label{Sect:Bssa}

Estimates of the magnetic field strength for individual components can be
obtained under the assumption that synchrotron self-absorption (SSA) is the
dominant process shaping the observed spectra. Following
\cite{1983ApJ...264..296M}
the strength of the magnetic field can then be calculated as
\begin{equation}
 B_{{\rm SSA}} = 10^{-5} b(a) \, \vartheta^{4} \nu_{\rm m}^{5} S_{\rm m}^{-2}
\left(
\dfrac{\delta}{1+z}
\right) ~~~~ (G),
\label{B_SSA}
\end{equation}
where:
\[
\begin{array}{lp{0.8\linewidth}}
 b(a),        & is a tabulated spectral index-dependent parameter
                \cite[cf.][Table 1]{1983ApJ...264..296M}\\
 \vartheta,   & the component size in mas \\
 \nu_{\rm m}, & the spectral turnover frequency in GHz, and\\
 S_{\rm m},   & the spectral turnover flux density in Jy.
\end{array}
\]

Given the strong dependencies of $B_{{\rm SSA}}$, mainly on $\nu_{\rm m}$ and
$\vartheta$, we note that measurement uncertainties of those quantities,
translate into large uncertainty of $B_{{\rm SSA}}$, as given by Eq.
\ref{B_SSA}. Nevertheless, we attempt the estimation for those components that
the turnover frequency and flux density are somewhat constrained, namely knot
C1 and the core. Even if a Doppler factor $\delta$, correction is not applied,
an apparent value along the jet can be obtained \citep{2005A&A...433..815B}.
Table \ref{table:Bfield} summarizes our findings.

C1 exhibits a strong magnetic field of ${\sim} 5.0\delta$\,G.
This large value could be an indication of a highly-magnetized
disturbance traveling along the jet. Its high apparent velocity of
$22.1\,c$ and high Doppler factor ($22.1 \le \delta \le 51.4$) hint
towards this scenario.

The characteristic spectral flatness of the core is traditionally attributed to
the superposition of a number of individual SSA components, each with a
distinctly peaked spectrum. The spectral shape of the core at epoch $2009.9$
(see Fig. \ref{compSpectraAll}) hints the presence of a SSA peak close to a
frequency of ${\sim} 43$\,GHz. Taking this as the turnover frequency
and with $S_{\rm m} = 2.3$\,Jy we obtain a SSA magnetic field of
$B_{{\rm SSA}} \sim 2\delta$\,mG.

\begin{table}
\caption{Estimated physical parameters for knot C1 and the core.} 
\label{table:Bfield}
\centering          
\begin{tabular}{l c c c c c c c}
\hline
\hline
Knot      & $\nu_{\rm m}$ & $S_{\rm m}$ & $\vartheta$  & $B_{\rm SSA} / \delta$ & $B_{\rm eq}$ & $\delta_{\rm eq}$ \\	    
          &    (GHz)      &    (Jy)     &   (mas)      &	 (G)		&   (G)        &		   \\	    
\hline\noalign{\medskip}
C1        &  $36.7$       &   $0.13$    &  $0.10$      &     $     5.0   $	&  $0.3$       &  ${\sim} 50$     \\   
Core      & ${\sim} 43$   &   $2.3$     &  $0.06$      &     $  0.002	 $ &  $1.2$       &  \dots 	   \\
\noalign{\smallskip}
\hline                  
\end{tabular}
\tablefoot{Columns from left to right:
(1) knot designation;
(2) turnover frequency;
(3) turnover flux density;
(4) deconvolved size;
(5) SSA magnetic field;
(6) minimum equipartition magnetic field and
(7) Doppler factor.}
\end{table}

Another constraint to the magnetic field can be set assuming equipartition
between the energy of radiating relativistic particles ($E_{e} \propto
B^{-1.5}$) and that of the magnetic field ($E_{B} \propto B^{2}$). The
equipartition magnetic field that minimizes the total energy content $E_{\rm
tot} = (1+k) E_{e}+E_{B}$ is \cite[see e.g.][]{2005A&A...433..815B}
\begin{equation}
 B_{\rm eq} = \left[ 4.5 \, (1+k)\, f(a, \nu_{\rm a}, \nu_{\rm b}) \,L \,R^{3})
\right]^{2/7}
\end{equation}
where:
\[
\begin{array}{lp{0.8\linewidth}}
 k, & is energy ratio between electrons and heavy particles, \\
 f, & is a tabulated function of $\alpha$, and upper and lower synchrotron
      frequency cutoffs, $\nu_{\rm a}$, $\nu_{\rm b}$ in\,GHz \\
 L, & is the synchrotron luminosity of the source given by $L=4 \pi
      D_{\rm L}^{2} \int_{\nu_{\rm a}}^{\nu_{\rm b}} S d\nu$, and \\
 R, & is the size of the knot in cm.
\end{array}
\]
For $k \sim 100$ and $f(-0.5, 10^{7}, 10^{11}) = 1.6 \times 10^{7}$, we
obtain the following expression
\begin{equation}
 B_{\rm eq} = 5.37 \times 10^{12} \,(S_{\rm m} \nu_{\rm m} D_{\rm L}^{2}
R^{-3})^{2/7}.
\end{equation}

Knot C1 is characterized by an equipartition magnetic field of
about $0.3$\,G and the core features the highest equipartition magnetic field,
$B_{\rm eq} \sim 1.2$\,G. Results are shown in Table \ref{table:Bfield}.

The two magnetic field estimates have a different dependence on the Doppler
factor. On the one hand, $B_{\rm SSA} \propto \delta$ for the magnetic field
strength calculated from SSA, and $B_{\rm eq} \propto \delta^{(2/7\alpha +1)}$
for the minimum, equipartition magnetic field. We use those to get an estimate
of $\delta_{\rm eq}$, when possible, from the ratio $B_{\rm eq} / B_{\rm SSA} =
\delta_{\rm eq}^{2/7 \alpha}$. The estimated equipartition Doppler factor for
C1, $\delta_{\rm eq,\,C1} \sim 50$, is also in good agreement with
$\delta_{\rm var} \sim 51.4$. Given the assumptions, all values are in
reasonable agreement.

%
\section{Discussion}

\subsection{The $\gamma$-ray/radio flare}

The \textit{Fermi}/LAT light curve, adopted from \cite{2014MNRAS.441.1899F} and
shown in Fig. \ref{LCs_all} (top panel), spans the time  period between
MJD\,$54707$ (2008.66) through MJD\,$55911$ (2011.96). The source is already at
high state, since the beginning of our monitoring period. From then on,
the rising trend continues until MJD\,$54875$ (2009.12), when the absolute
maximum in the monthly averaged $\gamma$-ray photon flux (of ${\sim} 1.3 \times
10^{-6} \, {\rm cm^{-2}\,s^{-1}}$), is reached. In fact, there exists some
amount of sub-structure in the $\gamma$-ray light curve. A second -- this time
local -- maximum is reached on MJD\,$54959$ (2009.35) with comparable flux.
Altogether, activity at high energies, from its start until the first -- and
absolute -- minimum is reached on MJD\,$55295$ (2010.27), lasts for almost 650
days (shaded area in Fig. \ref{LCs_all}).

In \cite{2002A&A...394..851S}, a general connection between radio
outbursts and newly ejected jet features has been made, with most radio flares
accompanied by the ejection of one or more components. While usually the
brightness of knots associated with $\gamma$-ray/radio flares decays fast, here
we follow the evolution of one thanks to the enhanced resolution of mm-VLBI. As
discussed in Sect. \ref{Sect:Decomp} apart from the flaring VLBI core, component
C3 visible at 43 and 86\,GHz only, is characterized by a decaying light curve
with an initial flux density level -- at epoch 2009.35 -- almost 10 times higher
than that of the other two components (C1, C2) present in the flow and seen also
at both high frequencies.
Although a certain amount of flux density blending between the core and
C3 is plaussible, we note the relatively high spatial separation of the two
components (approximately one beam at 86\,GHz in 2009.35) and the high
signal-to-noise ratio of C3. These two facts add to the argument that C3 is
indeed flaring/decaying, in addition to the flaring core.

Our observations are consistent with C3 having an apparent speed of $7.1\,c$
(Sect. \ref{kinematics}) and its ejection year, based on the kinematical
models at 43\,GHz and 86\,GHz, is ($2007.4 \pm 2.2$) and ($2008.3 \pm 0.3$),
respectively. C3's time of zero separation (red solid line in Fig.
\ref{LCs_all}) coincides with the onset of the $\gamma$-ray flare when
extrapolating the rate of $\gamma$-ray flux change, as it rises linearly, back
in time. We conclude, based on the ejection date of superluminal component C3,
that it is most probably associated with the $\gamma$-ray flare and its radio
counterpart. Since its separation from the core, it is traveling downstream the
jet with an apparent superluminal velocity (see Table \ref{86gigkin}). We note
the larger uncertainty in the knot's time of ejection at 43\,GHz as compared to
observation at 86\,GHz. That is due to the sparser sampling at this frequency
and the higher impact to the fit due to the very last data point in the radial
separation plot (middle panel of Fig. \ref{sepAll}). In any case, the two
figures are consistent with each other.
Assuming the speed of knot C3 to remain constant, at the times of the two
$\gamma$-ray photon flux maxima, in 2009.2 and 2009.4 (at MJD 54897 and 54959)
the knot is at a projected distance of ${\sim} 0.7$ and ${\sim}0.8$\,pc
downstream the 3-mm core, respectively.

A plausible scenario is that a disturbance originating at the jet nozzle of PKS
1502+106, while still upstream of the 3 mm core -- i.e. the unit-opacity surface
at 86\,GHz -- produces the observed $\gamma$-ray emission. As it continues to
move outwards, it becomes optically thin while crossing the 3-mm VLBI core hence
producing the radio flare. Alternatively, the core may represent the first
conical recollimation shock of the flow \citep{1988ApJ...334..539D,
1995ApJ...449L..19G, 2005MNRAS.357..918B, 2008ASPC..386..437M}, in which case
the flare can be attributed to shock--shock interaction and subsequent
enhancement of emission. In both cases, as the disturbance moves
further downstream, is observed as knot C3 at its decaying flux density phase.

\subsection{Distance estimates to the
central engine}\label{Sect:DistanceEstimates}

The debate as to where in the jet the high-energy emission
originates, is presently far from being considered resolved. In this paper,
combining radio single-dish monitoring data and our findings from the VLBI
analysis of PKS\,1502+106, we explore and further constrain the
region where the pronounced MeV/GeV activity of 2009 takes place and compare
with the SED modeling results of \cite{2010ApJ...710..810A}.

Under the assumption that the core takes up the entire jet cross section
and for a constant opening angle, the de-projected distance of the
86\,GHz core from the vertex of the hypothesized conical jet can be estimated
using the following expression
\begin{equation}
 d_{\rm core} = \dfrac{1.8\,\left< \mbox{FWHM}_{\rm core} \right>  }{2 \, \tan
(\phi_{\rm int}/2)}.
\label{eq:vlbi_d_core}
\end{equation}
With an average size for the core $\left< \mbox{FWHM}_{\rm core} \right>  =
0.03$\,mas and a nominal opening angle $\phi_{\rm int} = (3.8 \pm 0.5)^{\circ}$
(Sect. \ref{Sect:opAng}), the 86\,GHz core is
constrained to a distance of $d_{\rm core} \leq 8$\,pc away from
the jet base. This figure takes into account the uncertainty
in $\phi_{\rm int}$. Here, $d_{\rm core}$ is an upper limit, as the core is
at all times unresolved, rendering $\mbox{FWHM}_{\rm core}$ an upper limit
to its angular size.

In an effort to detect and assess possible correlations between radio and
$\gamma$ rays, \cite{2014MNRAS.441.1899F} apply a cross-correlation function
analysis between the single-dish radio light curves at 86\,GHz, and the
\textit{Fermi} $\gamma$-ray light curves of 54 blazars from the F-GAMMA sample.
Radio emission is typically found to lag behind the activity in $\gamma$ rays.
More specifically, for PKS\,1502+106, statistically significant correlation is
found between the two bands (see light curves in Fig. \ref{LCs_all}). Radio lags
$\gamma$ rays by $14 \pm 11$\,days at a significance level above 99\%.
Using Eq. 3 in the same paper, this time lag translates into a
distance of $2.1$\,pc upstream of the $\tau = 1$ surface at 86\,GHz, for
the $\gamma$-ray emission region.
Assuming now, that the core in our 86\,GHz images coincides
with the same surface of opacity transition from the optically thick to thin
regime, we conclude that the $\gamma$-ray emitting region is located at
${\leq}$5.9\,pc or ${\leq} 1.82 \times 10^{19}$\,cm away from the base of the
conical jet. These figures also represent upper limits for the distance between
the black hole and the $\gamma$-ray production region, due to the uncertain --
but likely small -- separation between the black hole horizon and the base of
the jet.

From the \ion{Mg}{ii} line profile, the bulk BLR radius of PKS\,1502+106 is
estimated to be $R_{\rm BLR} = 2 \times 10^{17}$\,cm \citep[see][and references
therein]{2010ApJ...710..810A}. Assuming a stratified and extended BLR, a
distance of $1.82 \times 10^{19}$\,cm corresponds to a region at its outer edge
or further away from it. The upper limit reported here agrees with the findings
of \cite{2010ApJ...710..810A}, where the external radiation Compton
mechanism (ERC) with the BLR photon field as a potential target for IC
up-scattering, has a significant contribution to the high-energy part of the
SED.
However, one should note the sharp drop of the BLR photon number density with
distance along with the fact that the $\gamma$-ray flare discussed in
\cite{2010ApJ...710..810A} precedes by a few weeks the one analyzed here, which
most likely belongs to the pre-main-flare period.
Consequently, the contribution of other seed photon fields (e.g. that
of the dusty torus, or other) must be explored.

The precision we can obtain from the VLBI data at hand is not sufficient for the
estimation of the magnetic field and other physical parameters, using
the core-shift effect \citep{1984ApJ...276...56M, 1998A&A...330...79L}. It is
worth noting that the results of \cite{2012A&A...545A.113P} place the 15\,GHz
core at a distance of ${\sim} 8$\,pc from the vertex of the jet. Since the
unit-opacity surface at 86\,GHz ought to be closer to the jet base
than the core at 15\,GHz, the $\gamma$-ray emission site is likely somewhat
closer to the jet base than our upper limit of 5.9\,pc suggests.
Finally, we note that Eq. \ref{eq:vlbi_d_core} must be used with caution in
estimating the distance between the core and the jet base through VLBI; being
the core unresolved, it yields only an upper limit to $d_{\rm core}$.

\subsection{Intrinsic jet properties}

\begin{figure}
 \centering
 \includegraphics[width=\linewidth]{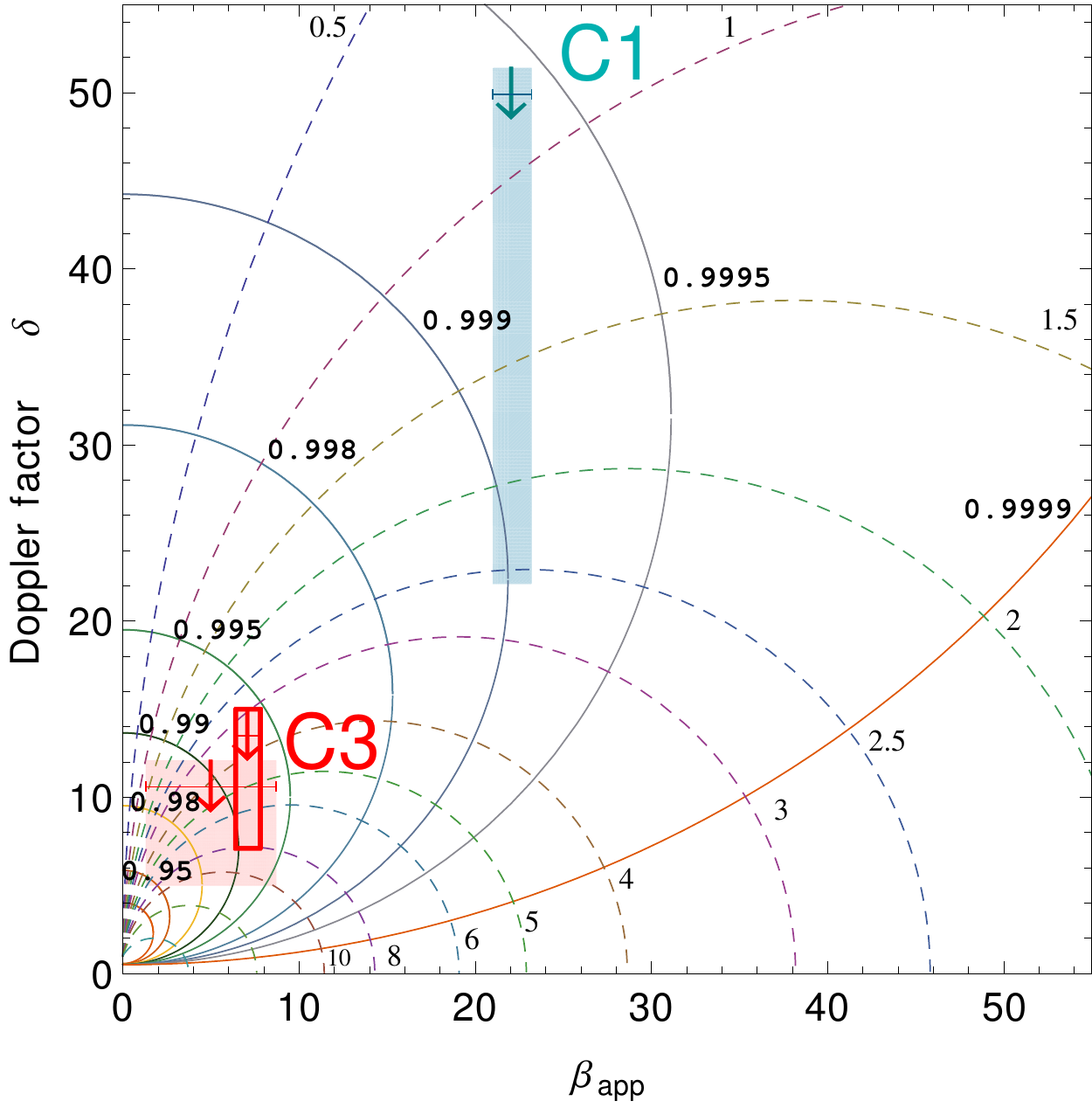}
  \caption{Doppler factor as a function of apparent speed for given viewing
angle, $\delta = f (\beta_{\rm app} | \theta)$, and given intrinsic speed
$\beta$, $\delta = f(\beta_{\rm app} | \beta)$. Loci represent these
two functions for a range of $\theta$ and $\beta$. The values of intrinsic
speeds plotted are $\beta = 0.9$, $0.95$, $0.98$, $0.99$, $0.995$, $0.998$,
$0.999$, $0.9995$, and $0.9999$ shown as solid lines.
For the viewing angle the values are $\theta =0.5^{\circ}$, $1^{\circ}$,
$1.5^{\circ}$, $2^{\circ}$, $2.5^{\circ}$, $3^{\circ}$, $4^{\circ}$,
$5^{\circ}$, $6^{\circ}$, $8^{\circ}$, $10^{\circ}$, $15^{\circ}$, and
$30^{\circ}$ shown as dashed lines. Note the two different jet regions, in
terms of physical characteristics, where components C1 and C3 are traveling.
Shaded areas represent the observables from our VLBI monitoring, namely the
Doppler factor and the apparent velocity for each of the two knots.}
 \label{gamato}%
\end{figure}

Given the Doppler factor and apparent speed of superluminal
knots C1 and C3, estimated in Sect. \ref{kinematics} and \ref{Sect:Dfctr}, we
explore the jet properties of PKS\,1502+106. From the expressions of both
the Doppler factor and Lorentz factor and making use of Eq. \ref{apparent}, we
obtain the expressions for the Doppler factor as a function of apparent speed
given the viewing angle, $\delta = f (\beta_{\rm app} | \theta)$, and given the
Lorentz factor, $\delta = f(\beta_{\rm app} | \gamma)$. The loci in Fig.
\ref{gamato} represent these two functions for a range of viewing angles and
Lorentz factors (see caption for details). Overplotted are regions of estimated
$\delta$ and $\beta_{\rm app}$. These help us in constraining the intrinsic
properties of the flow in terms of the viewing angle and the Lorentz factor.

Component C1 is characterized by an extreme variability Doppler factor of
${\sim} 51$. This figure is calculated based on causality arguments and agrees
well with $\delta_{\rm eq}$ (see Tables \ref{table:4} and \ref{table:Bfield}).
The same variability analysis for knot C3, suggests comparable results from both
43 and 86\,GHz, with Doppler factor ${\sim} 12$--$15$ (Table \ref{table:4}).
Fig. \ref{gamato}, where the inferred ranges of Doppler factor and apparent
speed are shown, is indicative of the two regions present in the relativistic
jet flow.

The difference in the inferred aspect angle and intrinsic speed, between
these two knots, reflect physical differences in the jet flow of
PKS\,1502+106. First, the jet we observe at high frequencies -- knot C3 -- is
possibly still in its accelerating phase (see also the apparent speed profile in
Fig. \ref{beta_appa}). This could explain the aforementioned differences (in
$\theta$ and $\delta$) seen between knots C3 and C1, with the latter being some
1\,mas downstream of the core. The jet, as revealed from the maps and the
overall structure, appears to bend, hence the high Doppler factor further out,
traced by knot C1, could also be attributed to differential Doppler boosting.
Taking into account only the $86$\,GHz data (red rectangle in Fig. \ref{gamato})
the two regions wherein components C3 and C1 are traveling are completely
incompatible, when it comes to viewing angle and intrinsic speed.

The difference between the Doppler boosting factors in two regions, at two
different viewing angles $\theta_{1}$ and $\theta_{2}$, keeping $\beta$ constant
is given by
\begin{equation}
 \delta_{2} - \delta_{1} = 
 \dfrac{ \sqrt{1-\beta^{2}} (\beta \cos \theta_{2} - \beta \cos \theta_{1})
}{(1- \beta \cos \theta_{2}) (1 - \beta \cos \theta_{1})}.
\label{eq:DiffDoppler}
\end{equation}
Here, we test whether the change of Doppler factor of ${\sim} 40$, between the
inner and outer jet, at distances ${<} 0.5$\,mas and ${>} 1$\,mas can be caused
only by the jet bending towards our line of sight. Components C3 and C1 are
traveling in those two regions, respectively. The Doppler factor is estimated
from variability -- for both -- and also from comparing the SSA and
equipartition magnetic field for C1. C3 is characterized by $\delta_{\rm var}
\sim 12$--$15$ while the two independent estimates agree very well for the
Doppler factor of C1 to an extreme value of ${\sim} 50$. From Fig. \ref{gamato}
the intrinsic speed of C3 is constrained between $\beta \sim 0.95$ and $\beta
\sim 0.995$. Taking even the highest value for $\beta$, keeping it constant, and
applying Eq. \ref{eq:DiffDoppler} between $\theta_{1} = 0.8^{\circ}$ and
$\theta_{2} = 3.4^{\circ}$ (the maximum range observed), we obtain that the
difference in Doppler factor of about 40 between the two regions cannot be
reconciled only by changing the viewing angle. For these numbers only a
$\delta_{2} - \delta_{1} \sim 5$ can be expected.
Thus, the two regions are characterized by both a change in the viewing angle
and also by acceleration -- i.e. change of Lorentz factor. This can also be seen
directly from the plot in Fig. \ref{gamato}, since only a few loci can lead
from one region to the other and only for the less constrained region for C3,
coming from the kinematics at $43$\,GHz. Converting the aforementioned
intrinsic velocities to the more convenient Lorentz factor, we obtain that even
for the highest $\beta = 0.995$ (for C3) and for $\beta = 0.999$ (for C1), a
change of Lorentz factor from, at least, $\gamma \sim 10$ to
$\gamma \sim 22$ takes place in the region between roughly $0.3$ to $1.1$\,mas.
This large-scale acceleration region speaks in favor of magnetic acceleration
scenarios \cite[see e.g.][]{2007MNRAS.380...51K}.

In estimating the viewing angle, it is worth noting that the critical
value, $\theta_{\rm c}$, might not be a good estimate for
the angle between the jet and the observer's line of sight. This is the
case for small kinematical Lorentz factors ($\gamma_{\rm min}$), deduced by low
observed apparent speeds. In fact, the viewing angle can be larger than
$\theta_{\rm c}$ for a flow at a larger viewing angle but intrinsically faster.
However, the critical angle does not significantly differ from the true $\theta$
for ultra-relativistic flows. This conclusion draws from the behavior of
$\beta_{\rm app}$ with respect to the viewing angle (cf. any plot of $\beta_{\rm
app}$ vs $\theta$) and the following energetics argument.
An electron traveling at $0.6\,c$ has only $1.5$ times more (relativistic)
kinetic energy than an electron at $0.5\,c$. The same electron at $0.999\,c$ or
$\gamma \sim 22$, possesses $3.4$ times more energy as compared to an $e^{-}$
traveling at a speed of $0.99\,c$ or $\gamma \sim 7$. Obviously it is harder
to accelerate the electron in the latter case as opposed to the former, making
the minimum Lorentz factor and consequently the critical angle good proxies for
the intrinsic values when very high apparent speeds are observed.

The variability viewing angle $\theta_{\rm var} \sim 0.8^{\circ}$, calculated
for C1 seems small compared to the critical one $\theta_{\rm c} = 2.6^{\circ}$.
These are not inconsistent findings though, because the critical viewing angle
represents an upper limit as well, calculated from the fastest
component -- here being C1. Estimation of a variability viewing angle is also
possible combining single-dish flux density measurements and VLBI kinematics as
in \cite{2009A&A...494..527H}, where the authors estimate $\theta_{\rm var} =
4.7^{\circ}$ for PKS\,1502+106. However, this figure is based on a smaller
$\beta_{\rm app} = 14.6$ measured at 22\,GHz and 37\,GHz in the period from
1990.5 to 1995.5 \citep[cf.][]{1998A&AS..132..305T}.

\cite{2009A&A...507L..33P} calculate a slightly smaller intrinsic opening angle
of $3.11^{\circ}$ for the jet. The reason is twofold. First, they obtain a
slightly smaller intrinsic opening angle of $37.9^{\circ}$ which they
subsequently de-project using the variability viewing angle reported in
\cite{2009A&A...494..527H}, thus obtaining the aforementioned value. The
difference with the opening angle of $(3.8 \pm 0.5)^{\circ}$ we find here is
justified by the higher apparent speed of component C1 seen in our data,
traveling with a $\beta_{\rm app}$ of 22.1.

The radial brightness temperature distribution cannot corroborate nor
falsify the two-region scenario, since in the first mas beyond the core most
figures reported constitute lower limits. This arises from the components being
unresolved at all three frequencies. In any case, what is evident is the
decaying trend at beyond 1\,mas (see Fig. \ref{Tb2}).
The scenario of acceleration within the first mas from the core, supported by
both kinematics and variability, cannot thus be tested through the $T_{\rm b}$.
The case in which the part of the jet probed by the 86\,GHz VLBI data is still
at its accelerating phase \cite[see e.g.][]{2013JKAS...46..243L}, with the
Doppler factor increasing downstream of the 86\,GHz core, is not incompatible
with the radial $T_{\rm b}$ profile. Additionally, we note that there is no
evidence for any overall rising trend of $T_{\rm b}$ with distance as we move
away from the core, but rather a steady decay seen at all frequencies.

%
\section{Summary and conclusions}\label{sect:VLBIconcl}

In this paper a comprehensive, three-frequency ($15$, $43$, and
$86$\,GHz) VLBI study of the $\gamma$-ray blazar PKS $1502{+}106$ was presented,
using GMVA and additional data from the MOJAVE monitoring program. Furthermore,
we employed the densely-sampled F-GAMMA single-dish light curves at
matching frequencies, along with the \textit{Fermi}/LAT monthly-binned light
curve at energies ${>}100$\,MeV. The work presented here allows us to follow
the multi-frequency flare of $2008$--$2010$ and conclude the following:

\begin{enumerate}

\item PKS $1502{+}106$ exhibits a compact, core-dominated morphology at all
three frequencies with a one-sided, bent, parsec-scale jet. Component
motion within the flow is characterised by high superluminal speeds in the
range $5$--$22$\,$c$.
The most extreme example is component C1 traveling downstream the jet at about
$22\,c$ at $15$ GHz and comparable speeds at $43/86$\,GHz.

\item Doppler and Lorentz factor estimates for individual components
are obtained using variability and VLBI kinematical arguments. The kinematical
Doppler and Lorentz factors are $\delta_{\rm C1} \sim
\gamma_{\rm  min,\,C1} = 22.1$ at 15\,GHz. For knots C2 and C3, it is
$\delta_{\rm C2} \sim \gamma_{\rm  min,\,C2} = 9.3$ and
$\delta_{\rm C3} \sim \gamma_{\rm min,\,C3} = 7.2$, respectively.
From causality, knot C1, showing significant variability at
$15$\,GHz, is characterized by $\delta_{\rm var}$ as high as 51.4 and
$\gamma_{\rm var} \sim 30.5$.
Knot C3 is also found to have consistent behavior at both high frequencies with
$\delta_{\rm var}^{\rm 43\,GHz} \sim 12.1$ and $\gamma_{\rm var}^{\rm 43\,GHz}
\sim 7.1$, while at 86\,GHz the estimation yields $\delta_{\rm var}^{\rm
86\,GHz} \sim 15$ and $\gamma_{\rm var}^{\rm 86\,GHz} \sim 9.2$. All figures
are consistent with the estimates of the kinematical Doppler and Lorentz
factors. Doppler factors are strikingly different for knot C3 at $r < 0.5$\,mas
and C1 at a distance $r > 1$\,mas from the core.

\item An additional Doppler factor can be calculated from the different
dependencies of $B_{\rm SSA}$ and $B_{\rm eq}$ on $\delta$. The estimate for
C1 is in agreement with those from kinematics and causality.

\item Using variability arguments we are able to
constrain the viewing angle towards the source to $\theta \sim 3.4^{\circ}$ for
the inner and $\theta \sim 0.8^{\circ}$ for the outer portions of the jet, after
about $1$\,mas. The calculation of the critical viewing angle for knot C1 at
$15$\,GHz yields $\theta_{\rm c} = 2.6^{\circ}$.

\item We calculate the opening angle of the jet of PKS\,$1502{+}106$
with two methods with the results in very good agreement.
The nominal de-projected opening angle is $(3.8 \pm 0.5)^{\circ}$.

\item The differences in apparent speed and viewing angle point towards a
jet bending after the first mas. This results in differential Doppler
boosting -- i.e. increasing $\delta$ from ${\sim} 12$--$15$ to ${\sim} 50$, as
traced by components C3 and C1, traveling at radial distances of ${<} 0.5$\,mas
and  ${>} 1$\,mas from the core, respectively. However, this Doppler factor
gradient cannot be ascribed to jet bending only. Acceleration must
also be at play within the first mas of the jet.

\item By constraining the intrinsic speed, $\beta$, and viewing angle, $\theta$,
from the calculated Doppler factors and apparent velocities, we can clearly
distinguish between two regions of the jet, wherein knots C1 and C3 are
traveling. Specifically, closer to the core at a distance ${\leq} 0.5$\,mas the
parameters obtained for C3 at 43\,GHz constrain the parameters of the flow to
$0.95 \leq \beta \leq 0.995$ and $1.5^{\circ} \leq \theta \leq 10^{\circ}$. The
same component at 86\,GHz sets slightly stronger constraints to the viewing
angle with $3^{\circ} \leq \theta \leq 8^{\circ}$. On the other hand, in the
region where C1 is traveling, its observed $\delta$ and $\beta_{\rm app}$
constrain the flow to $\beta \geq 0.999$ and $0.8^{\circ} \leq \theta \leq
2.5^{\circ}$. A clear division is seen between these two jet regions.
Hence, we conclude that the jet of PKS $1502{+}106$ bends towards us in the
region beyond ${\sim} 1$\,mas downstream of the core while also accelerating.

\item The radial brightness temperature profile indicates very high $T_{\rm b}$
for the core region at all three frequencies.
We find though a trend of decreasing $T_{\rm b}$ as we go towards higher
observing frequency and with increasing distance from the core, with no
increasing trend. This could indicate a magnetically-dominated core
region probed by the $86$\,GHz observations where Poynting flux has not yet
fully converted to kinetic flux.

\item Component spectra allow the calculation of the magnetic field strength
in different regions of the jet. Component C1 exhibits a distinctive SSA peak.
The core shows a flat spectrum with $\left< \alpha \right> \sim -0.22$ but at
one epoch ($2009.9$) signs of a turnover at ${\sim} 43$\,GHz.
Through the comparison between the SSA and equipartition, minimum magnetic
fields we conclude that for component C1, $B_{\rm SSA}$ is much stronger than
$B_{\rm eq}$, indicating a high degree of Doppler boosting. Finally, for the
core it is $B_{\rm SSA}/\delta \ge 2$\,mG and $B_{\rm eq} = 1.2$\,G.

\item The radio flux density decomposition into distinct VLBI components and
comparison with the single-dish radio flux density outburst, indicate that the
bulk of radio emission originates from the core at all frequencies.
Apart from the core at $43/86$\,GHz, knot C3 shares a significant radio flux
density level which decays with time. C3 is not resolved at $15$\,GHz
due to blending with the core.

\item Given the radio flux density decomposition and the estimated ejection
time of knot C3 -- coincident with the single-dish mm flare onset -- we conclude
that it is responsible for the radio flare observed in PKS\,$1502{+}106$ during
the period $2008$--$2010$. Furthermore, the previously established correlation
between the radio flare and $\gamma$ rays indicates that component C3 is
responsible for the $\gamma$-ray flare and its radio counterparts. Arguably, the
$2008$--$2010$ flare of PKS\,$1502{+}106$ is an event originating in a disturbed
region at the jet nozzle. While still upstream of the $3$-mm core, it produces
the increased $\gamma$-ray emission observed and as it continues to move, it 
becomes optically thin (at $86$\,GHz) close to the $3$-mm VLBI core thus
enhancing its brightness. It then continues on and after the 86\,GHz optically
thick region, is seen as knot C3 at its decaying flux density phase.

\item We conclude that the $\gamma$-ray emitting region is located at
${\leq} 5.9$\,pc or ${\leq} 1.82 \times 10^{19}$\,cm away from the base of the
jet. These estimates constitute upper limits for the distance between the
$\gamma$-ray production region and the SMBH itself.
\end{enumerate}

Concluding, PKS $1502{+}106$ represents a source whose complex structural
dynamics needs to be further investigated with higher-cadence, high-resolution
imaging at mas and sub-mas scales.

\begin{longtab}
\small
\begin{longtable}{c c c c c c c c}
\caption{PKS\,1502+106 \texttt{MODELFIT} results at 15\,GHz.}\\
\hline\hline
Epoch    &   MJD     & Epoch &  S     &   r    &	   PA	   &  FWHM	  &   ID\\
         &           &   ID  & (Jy)   & (mas)  & ($^{\circ}$) &  (mas)  &     \\
\hline
\endfirsthead
\caption{continued.}\\
\hline\hline
Epoch    &   MJD    & Epoch &     S     &   r    &      PA      &   FWHM  &   ID\\
         &          &	ID  &    (Jy)   & (mas)  & ($^{\circ}$) &  (mas)  &     \\
\hline
\endhead
\hline
\endfoot
\noalign{\medskip}
2009.23  &  54915.5   &  13  &  $0.038 \pm 0.004$  &  $4.61 \pm 0.13$	&  $103 \pm  11$  &  $4.57 \pm 0.46$   &   \dots   \\
\noalign{\medskip}
2002.61  &  52498.5   &  1   &  $0.110 \pm 0.011$  &  $1.98 \pm 0.28$	&  $123 \pm  8 $  &  $1.38 \pm 0.14$   &   Ca  \\
2003.24  &  52727.5   &  2   &  $0.103 \pm 0.010$  &  $2.30 \pm 0.32$	&  $123 \pm  8 $  &  $1.60 \pm 0.16$   &   Ca  \\
2004.80  &  53296.5   &  3   &  $0.114 \pm 0.011$  &  $2.03 \pm 0.36$	&  $126 \pm  10$  &  $1.79 \pm 0.18$   &   Ca  \\
2005.36  &  53503.5   &  4   &  $0.079 \pm 0.008$  &  $2.65 \pm 0.37$	&  $123 \pm  8 $  &  $1.83 \pm 0.18$   &   Ca  \\
2005.73  &  53636.5   &  5   &  $0.102 \pm 0.010$  &  $2.47 \pm 0.38$	&  $122 \pm  9 $  &  $1.91 \pm 0.19$   &   Ca  \\
2005.83  &  53672.5   &  6   &  $0.081 \pm 0.008$  &  $2.60 \pm 0.31$	&  $122 \pm  7 $  &  $1.55 \pm 0.16$   &   Ca  \\
2005.88  &  53691.5   &  7   &  $0.096 \pm 0.010$  &  $2.27 \pm 0.25$	&  $122 \pm  6 $  &  $1.23 \pm 0.12$   &   Ca  \\
2006.52  &  53923.5   &  8   &  $0.093 \pm 0.009$  &  $2.38 \pm 0.27$	&  $123 \pm  7 $  &  $1.36 \pm 0.14$   &   Ca  \\
2007.62  &  54328.5   &  9   &  $0.118 \pm 0.012$  &  $2.59 \pm 0.33$	&  $123 \pm  7 $  &  $1.66 \pm 0.17$   &   Ca  \\
2008.48  &  54642.5   &  10  &  $0.134 \pm 0.013$  &  $2.46 \pm 0.42$	&  $123 \pm  10$  &  $2.09 \pm 0.21$   &   Ca  \\
2008.60  &  54684.5   &  11  &  $0.100 \pm 0.010$  &  $2.84 \pm 0.36$	&  $123 \pm  7 $  &  $1.78 \pm 0.18$   &   Ca  \\
2008.88  &  54789.5   &  12  &  $0.112 \pm 0.011$  &  $2.79 \pm 0.42$	&  $123 \pm  9 $  &  $2.11 \pm 0.21$   &   Ca  \\
2009.23  &  54915.5   &  13  &  $0.064 \pm 0.006$  &  $2.93 \pm 0.30$	&  $125 \pm  6 $  &  $1.48 \pm 0.15$   &   Ca  \\
2009.94  &  55175.5   &  14  &  $0.088 \pm 0.009$  &  $2.97 \pm 0.46$	&  $122 \pm  9 $  &  $2.30 \pm 0.23$   &   Ca  \\
2010.47  &  55366.5   &  15  &  $0.094 \pm 0.009$  &  $3.04 \pm 0.53$	&  $121 \pm  10$  &  $2.67 \pm 0.27$   &   Ca  \\
2010.65  &  55435.5   &  16  &  $0.096 \pm 0.010$  &  $3.00 \pm 0.54$	&  $121 \pm  10$  &  $2.70 \pm 0.27$   &   Ca  \\
2010.87  &  55513.5   &  17  &  $0.098 \pm 0.010$  &  $3.02 \pm 0.58$	&  $121 \pm  11$  &  $2.91 \pm 0.29$   &   Ca  \\
2011.16  &  55619.5   &  18  &  $0.090 \pm 0.009$  &  $3.08 \pm 0.58$	&  $120 \pm  11$  &  $2.89 \pm 0.29$   &   Ca  \\
2011.62  &  55788.5   &  19  &  $0.098 \pm 0.010$  &  $3.22 \pm 0.67$	&  $119 \pm  12$  &  $3.35 \pm 0.33$   &   Ca  \\
\noalign{\medskip}
2002.61  &  52498.5   &  1   &  $0.165 \pm 0.017$  &  $0.68 \pm 0.07$	&  $120 \pm  6 $  &  $0.32 \pm 0.03$   &   Cb  \\
2003.24  &  52727.5   &  2   &  $0.155 \pm 0.016$  &  $0.97 \pm 0.11$	&  $121 \pm  6 $  &  $0.54 \pm 0.05$   &   Cb  \\
2004.80  &  53296.5   &  3   &  $0.141 \pm 0.014$  &  $0.84 \pm 0.13$	&  $117 \pm  9 $  &  $0.65 \pm 0.07$   &   Cb  \\
2005.36  &  53503.5   &  4   &  $0.127 \pm 0.013$  &  $1.16 \pm 0.21$	&  $122 \pm  10$  &  $1.03 \pm 0.10$   &   Cb  \\
2005.73  &  53636.5   &  5   &  $0.096 \pm 0.010$  &  $1.21 \pm 0.19$	&  $124 \pm  9 $  &  $0.96 \pm 0.10$   &   Cb  \\
2005.83  &  53672.5   &  6   &  $0.108 \pm 0.011$  &  $1.19 \pm 0.19$	&  $121 \pm  9 $  &  $0.97 \pm 0.10$   &   Cb  \\
2005.88  &  53691.5   &  7   &  $0.045 \pm 0.005$  &  $1.26 \pm 0.10$	&  $136 \pm  4 $  &  $0.49 \pm 0.05$   &   Cb  \\
2006.52  &  53923.5   &  8   &  $0.033 \pm 0.003$  &  $1.19 \pm 0.14$	&  $141 \pm  7 $  &  $0.10 \pm 0.01$   &   Cb  \\
\noalign{\medskip}
2003.24  &  52727.5   &  2   &  $0.101 \pm 0.010$  &  $0.37 \pm 0.13$	&  $123 \pm  19$  &  $0.10 \pm 0.01$   &   Cc  \\
2004.80  &  53296.5   &  3   &  $0.129 \pm 0.013$  &  $0.42 \pm 0.13$	&  $98  \pm  17$  &  $0.10 \pm 0.01$   &   Cc  \\
2005.36  &  53503.5   &  4   &  $0.157 \pm 0.016$  &  $0.58 \pm 0.10$	&  $101 \pm  10$  &  $0.50 \pm 0.05$   &   Cc  \\
2005.73  &  53636.5   &  5   &  $0.119 \pm 0.012$  &  $0.64 \pm 0.10$	&  $100 \pm  9 $  &  $0.52 \pm 0.05$   &   Cc  \\
2005.83  &  53672.5   &  6   &  $0.087 \pm 0.009$  &  $0.66 \pm 0.10$	&  $99  \pm  8 $  &  $0.48 \pm 0.05$   &   Cc  \\
2005.88  &  53691.5   &  7   &  $0.110 \pm 0.011$  &  $0.71 \pm 0.12$	&  $97  \pm  9 $  &  $0.57 \pm 0.06$   &   Cc  \\
2006.52  &  53923.5   &  8   &  $0.061 \pm 0.006$  &  $0.82 \pm 0.15$	&  $90  \pm  10$  &  $0.75 \pm 0.07$   &   Cc  \\
2008.48  &  54642.5   &  10  &  $0.037 \pm 0.004$  &  $1.30 \pm 0.13$	&  $99  \pm  6 $  &  $0.10 \pm 0.01$   &   Cc  \\
2008.60  &  54684.5   &  11  &  $0.101 \pm 0.010$  &  $1.14 \pm 0.14$	&  $103 \pm  7 $  &  $0.72 \pm 0.07$   &   Cc  \\
2008.88  &  54789.5   &  12  &  $0.036 \pm 0.004$  &  $1.47 \pm 0.08$	&  $103 \pm  3 $  &  $0.39 \pm 0.04$   &   Cc  \\
2009.23  &  54915.5   &  13  &  $0.056 \pm 0.006$  &  $1.35 \pm 0.20$	&  $105 \pm  8 $  &  $0.98 \pm 0.10$   &   Cc  \\
\noalign{\medskip}
2007.62  &  54328.5   &  9   &  $0.278 \pm 0.028$  &  $0.61 \pm 0.08$	&  $97  \pm  7 $  &  $0.38 \pm 0.04$   &   C1  \\
2008.48  &  54642.5   &  10  &  $0.118 \pm 0.012$  &  $0.77 \pm 0.05$	&  $94  \pm  3 $  &  $0.23 \pm 0.02$   &   C1  \\
2008.60  &  54684.5   &  11  &  $0.072 \pm 0.007$  &  $0.76 \pm 0.14$	&  $91  \pm  11$  &  $0.10 \pm 0.01$   &   C1  \\
2008.88  &  54789.5   &  12  &  $0.147 \pm 0.015$  &  $0.80 \pm 0.06$	&  $95  \pm  5 $  &  $0.32 \pm 0.03$   &   C1  \\
2009.23  &  54915.5   &  13  &  $0.054 \pm 0.005$  &  $0.92 \pm 0.15$	&  $92  \pm  9 $  &  $0.10 \pm 0.01$   &   C1  \\
2009.94  &  55175.5   &  14  &  $0.037 \pm 0.004$  &  $1.26 \pm 0.18$	&  $98  \pm  8 $  &  $0.89 \pm 0.09$   &   C1  \\
2010.47  &  55366.5   &  15  &  $0.037 \pm 0.004$  &  $1.30 \pm 0.15$	&  $97  \pm  7 $  &  $0.77 \pm 0.08$   &   C1  \\
2010.65  &  55435.5   &  16  &  $0.027 \pm 0.003$  &  $1.38 \pm 0.13$	&  $93  \pm  5 $  &  $0.64 \pm 0.06$   &   C1  \\
2010.87  &  55513.5   &  17  &  $0.026 \pm 0.003$  &  $1.47 \pm 0.13$	&  $96  \pm  5 $  &  $0.66 \pm 0.07$   &   C1  \\
2011.16  &  55619.5   &  18  &  $0.020 \pm 0.002$  &  $1.57 \pm 0.09$	&  $98  \pm  3 $  &  $0.43 \pm 0.04$   &   C1  \\
2011.62  &  55788.5   &  19  &  $0.027 \pm 0.003$  &  $1.58 \pm 0.12$	&  $105 \pm  4 $  &  $0.59 \pm 0.06$   &   C1  \\
\noalign{\medskip}
2008.48  &  54642.5   &  10  &  $0.105 \pm 0.011$  &  $0.32 \pm 0.13$	&  $90  \pm  23$  &  $0.10 \pm 0.01$   &   C2  \\
2008.60  &  54684.5   &  11  &  $0.117 \pm 0.012$  &  $0.35 \pm 0.14$	&  $98  \pm  22$  &  $0.10 \pm 0.01$   &   C2  \\
2008.88  &  54789.5   &  12  &  $0.100 \pm 0.010$  &  $0.43 \pm 0.14$	&  $101 \pm  19$  &  $0.10 \pm 0.01$   &   C2  \\
2009.23  &  54915.5   &  13  &  $0.137 \pm 0.014$  &  $0.58 \pm 0.06$	&  $99  \pm  6 $  &  $0.32 \pm 0.03$   &   C2  \\
2009.94  &  55175.5   &  14  &  $0.141 \pm 0.014$  &  $0.74 \pm 0.09$	&  $100 \pm  7 $  &  $0.44 \pm 0.04$   &   C2  \\
2010.47  &  55366.5   &  15  &  $0.123 \pm 0.012$  &  $0.71 \pm 0.11$	&  $103 \pm  9 $  &  $0.54 \pm 0.05$   &   C2  \\
2010.65  &  55435.5   &  16  &  $0.120 \pm 0.012$  &  $0.74 \pm 0.11$	&  $103 \pm  9 $  &  $0.57 \pm 0.06$   &   C2  \\
2010.87  &  55513.5   &  17  &  $0.112 \pm 0.011$  &  $0.77 \pm 0.11$	&  $105 \pm  8 $  &  $0.56 \pm 0.06$   &   C2  \\
2011.16  &  55619.5   &  18  &  $0.118 \pm 0.012$  &  $0.81 \pm 0.11$	&  $108 \pm  8 $  &  $0.56 \pm 0.06$   &   C2  \\
2011.62  &  55788.5   &  19  &  $0.098 \pm 0.010$  &  $0.82 \pm 0.14$	&  $113 \pm  10$  &  $0.69 \pm 0.07$   &   C2  \\
\noalign{\medskip}
2002.61  &  52498.5   &  1   &  $1.341 \pm 0.134$  &  $0.00 \pm 0.02$	&  $0.0    \pm  0.0 $  &  $0.23 \pm 0.02$   &	Core  \\
2003.24  &  52727.5   &  2   &  $1.457 \pm 0.146$  &  $0.00 \pm 0.02$	&  $0.0    \pm  0.0 $  &  $0.22 \pm 0.02$   &	Core  \\
2004.80  &  53296.5   &  3   &  $0.611 \pm 0.061$  &  $0.00 \pm 0.02$	&  $0.0    \pm  0.0 $  &  $0.20 \pm 0.02$   &	Core  \\
2005.36  &  53503.5   &  4   &  $0.505 \pm 0.051$  &  $0.00 \pm 0.12$	&  $0.0    \pm  0.0 $  &  $0.11 \pm 0.01$   &	Core  \\
2005.73  &  53636.5   &  5   &  $0.803 \pm 0.080$  &  $0.00 \pm 0.01$	&  $0.0    \pm  0.0 $  &  $0.12 \pm 0.01$   &	Core  \\
2005.83  &  53672.5   &  6   &  $0.868 \pm 0.087$  &  $0.00 \pm 0.12$	&  $0.0    \pm  0.0 $  &  $0.11 \pm 0.01$   &	Core  \\
2005.88  &  53691.5   &  7   &  $0.926 \pm 0.093$  &  $0.00 \pm 0.15$	&  $0.0    \pm  0.0 $  &  $0.11 \pm 0.01$   &	Core  \\
2006.52  &  53923.5   &  8   &  $1.339 \pm 0.134$  &  $0.00 \pm 0.02$	&  $0.0    \pm  0.0 $  &  $0.23 \pm 0.02$   &	Core  \\
2007.62  &  54328.5   &  9   &  $1.113 \pm 0.111$  &  $0.00 \pm 0.12$	&  $0.0    \pm  0.0 $  &  $0.09 \pm 0.01$   &	Core  \\
2008.48  &  54642.5   &  10  &  $1.348 \pm 0.135$  &  $0.00 \pm 0.12$	&  $0.0    \pm  0.0 $  &  $0.09 \pm 0.01$   &	Core  \\
2008.60  &  54684.5   &  11  &  $1.320 \pm 0.132$  &  $0.00 \pm 0.12$	&  $0.0    \pm  0.0 $  &  $0.08 \pm 0.01$   &	Core  \\
2008.88  &  54789.5   &  12  &  $1.589 \pm 0.159$  &  $0.00 \pm 0.12$	&  $0.0    \pm  0.0 $  &  $0.08 \pm 0.01$   &	Core  \\
2009.23  &  54915.5   &  13  &  $2.815 \pm 0.282$  &  $0.00 \pm 0.13$	&  $0.0    \pm  0.0 $  &  $0.06 \pm 0.01$   &	Core  \\
2009.94  &  55175.5   &  14  &  $1.213 \pm 0.121$  &  $0.00 \pm 0.12$	&  $0.0    \pm  0.0 $  &  $0.09 \pm 0.01$   &	Core  \\
2010.47  &  55366.5   &  15  &  $0.723 \pm 0.072$  &  $0.00 \pm 0.13$	&  $0.0    \pm  0.0 $  &  $0.10 \pm 0.01$   &	Core  \\
2010.65  &  55435.5   &  16  &  $0.749 \pm 0.075$  &  $0.00 \pm 0.12$	&  $0.0    \pm  0.0 $  &  $0.09 \pm 0.01$   &	Core  \\
2010.87  &  55513.5   &  17  &  $0.880 \pm 0.088$  &  $0.00 \pm 0.12$	&  $0.0    \pm  0.0 $  &  $0.09 \pm 0.01$   &	Core  \\
2011.16  &  55619.5   &  18  &  $1.003 \pm 0.100$  &  $0.00 \pm 0.12$	&  $0.0    \pm  0.0 $  &  $0.09 \pm 0.01$   &	Core  \\
2011.62  &  55788.5   &  19  &  $1.127 \pm 0.113$  &  $0.00 \pm 0.13$	&  $0.0    \pm  0.0 $  &  $0.09 \pm 0.01$   &	Core  \\
\noalign{\medskip}
\end{longtable}
\tablefoot{Columns from left to right:
(1) observing epoch in fractional year;
(2) MJD of the observing epoch;
(3) epoch identifier between $1$--$19$; 
(4) integrated component flux density;
(5) radial separation from the core;
(6) position angle;
(7) component size given as the FWHM of the major axis;
(8) component identification label.}
\end{longtab}

\begin{longtab}
\small
\begin{longtable}{c c c c c c c c}
\caption{PKS\,1502+106 \texttt{MODELFIT} results at 43\,GHz.}\\
\hline
\hline
Epoch    &   MJD     & Epoch  &    S     &   r    &      PA      &  FWHM	&   ID\\
         &           &   ID   &   (Jy)   & (mas)  & ($^{\circ}$) &  (mas)  &	 \\
\hline
\endfirsthead
\caption{continued.}\\
\hline
\hline
Epoch    &   MJD     & Epoch  &  S     &	r    &      PA      &	FWHM  &   ID\\
         &           &   ID   & (Jy)   & (mas)  & ($^{\circ}$) &  (mas)  &     \\
\hline
\endhead
\hline
\endfoot
\noalign{\medskip}
2009.35  &  54959.36  &   1  & $0.055  \pm  0.006$  &  $0.93  \pm  0.08$ &    $94  \pm 5 $    & $0.404  \pm  0.040$ &	C1    \\
2009.78  &  55117.82  &   2  & $0.109  \pm  0.011$  &  $0.87  \pm  0.08$ &    $99  \pm 5 $    & $0.404  \pm  0.040$ &	C1    \\
2010.34  &  55323.28  &   3  & $0.044  \pm  0.004$  &  $1.08  \pm  0.19$ &    $97  \pm 10$    & $0.965  \pm  0.096$ &	C1    \\
2011.35  &  55688.76  &   4  & $0.056  \pm  0.006$  &  $1.20  \pm  0.24$ &    $107 \pm 11$    & $1.211  \pm  0.121$ &	C1    \\
2011.77  &  55843.86  &   5  & $0.043  \pm  0.004$  &  $1.20  \pm  0.20$ &    $103 \pm 10$    & $1.017  \pm  0.102$ &	C1    \\
2012.38  &  56065.17  &   6  & $0.041  \pm  0.004$  &  $1.35  \pm  0.29$ &    $97  \pm 12$    & $1.467  \pm  0.147$ &	C1    \\
\noalign{\medskip}
2009.35  &  54959.36  &   1  & $0.059  \pm  0.006$  &  $0.57  \pm  0.04$ &    $100 \pm 4 $    & $0.202  \pm  0.020$ &	C2    \\
2009.78  &  55117.82  &   2  & $0.044  \pm  0.004$  &  $0.49  \pm  0.07$ &    $99  \pm 8 $    & $0.040  \pm  0.004$ &	C2    \\
2010.34  &  55323.28  &   3  & $0.031  \pm  0.003$  &  $0.74  \pm  0.03$ &    $100 \pm 3 $    & $0.160  \pm  0.016$ &	C2    \\
2011.35  &  55688.76  &   4  & $0.025  \pm  0.003$  &  $0.78  \pm  0.04$ &    $108 \pm 3 $    & $0.191  \pm  0.019$ &	C2    \\
2011.77  &  55843.86  &   5  & $0.014  \pm  0.001$  &  $0.71  \pm  0.03$ &    $108 \pm 2 $    & $0.130  \pm  0.013$ &	C2    \\
2012.38  &  56065.17  &   6  & $0.011  \pm  0.001$  &  $0.65  \pm  0.03$ &    $120 \pm 3 $    & $0.159  \pm  0.016$ &	C2    \\
\noalign{\medskip}
2009.35  &  54959.36  &   1  & $0.554  \pm  0.055$  &  $0.08  \pm  0.06$ &    $130 \pm 37$    & $0.030  \pm  0.003$ &	C3    \\
2009.78  &  55117.82  &   2  & $0.108  \pm  0.011$  &  $0.14  \pm  0.07$ &    $126 \pm 26$    & $0.040  \pm  0.004$ &	C3    \\
2010.34  &  55323.28  &   3  & $0.042  \pm  0.004$  &  $0.22  \pm  0.04$ &    $115 \pm 10$    & $0.194  \pm  0.019$ &	C3    \\
2011.35  &  55688.76  &   4  & $0.032  \pm  0.003$  &  $0.18  \pm  0.08$ &    $61  \pm 23$    & $0.040  \pm  0.004$ &	C3    \\
\noalign{\medskip}
2011.77  &  55843.86  &   5  & $0.080  \pm  0.008$  &  $0.14  \pm  0.02$ &    $70  \pm 7 $    & $0.089  \pm  0.009$ &	\dots \\
2012.38  &  56065.17  &   6  & $0.068  \pm  0.007$  &  $0.13  \pm  0.07$ &    $70  \pm 28$    & $0.050  \pm  0.005$ &	\dots \\
\noalign{\medskip}
2009.35  &  54959.36  &   1  & $3.030  \pm  0.303$  &  $0.00  \pm  0.06$ &  $0.0   \pm 0.0 $  & $0.029  \pm  0.003$ &	Core  \\
2009.78  &  55117.82  &   2  & $2.301  \pm  0.230$  &  $0.00  \pm  0.07$ &  $0.0   \pm 0.0 $  & $0.056  \pm  0.006$ &	Core  \\
2010.34  &  55323.28  &   3  & $0.530  \pm  0.053$  &  $0.00  \pm  0.07$ &  $0.0   \pm 0.0 $  & $0.054  \pm  0.005$ &	Core  \\
2011.35  &  55688.76  &   4  & $1.067  \pm  0.107$  &  $0.00  \pm  0.03$ &  $0.0   \pm 0.0 $  & $0.131  \pm  0.013$ &	Core  \\
2011.77  &  55843.86  &   5  & $0.993  \pm  0.099$  &  $0.00  \pm  0.01$ &  $0.0   \pm 0.0 $  & $0.073  \pm  0.007$ &	Core  \\
2012.38  &  56065.17  &   6  & $0.799  \pm  0.080$  &  $0.00  \pm  0.07$ &  $0.0   \pm 0.0 $  & $0.040  \pm  0.004$ &	Core  \\
\noalign{\medskip}
\end{longtable}
\tablefoot{Columns from left to right:
(1) observing epoch in fractional year;
(2) MJD of the observing epoch;
(3) epoch identifier between $1$--$6$; 
(4) integrated component flux density;
(5) radial separation from the core;
(6) position angle;
(7) component size given as the FWHM of the major axis;
(8) component identification label.}
\end{longtab}

\begin{longtab}
\small
\begin{longtable}{c c c c c c c c}
\caption{PKS\,1502+106 \texttt{MODELFIT} results at 86\,GHz.}\\
\hline
\hline
Epoch    &   MJD     & Epoch  &    S     &   r    &      PA      &  FWHM	&   ID\\
         &           &   ID   &   (Jy)   & (mas)  & ($^{\circ}$) &  (mas)  &	 \\
\hline
\endfirsthead
\caption{continued.}\\
\hline
\hline
Epoch    &   MJD     & Epoch  &  S     &	r    &      PA      &	FWHM  &   ID\\
         &           &   ID   & (Jy)   & (mas)  & ($^{\circ}$) &  (mas)  &     \\
\hline
\endhead
\hline
\endfoot
\noalign{\medskip}
2009.35  &  54959.36 &  1  & $0.071 \pm 0.007$ & $1.18 \pm 0.16$  & $99  \pm \dots$  &  $0.780 \pm 0.078$  &  \dots \\
2010.34  &  55323.28 &  3  & $0.011 \pm 0.001$ & $1.39 \pm 0.08$  & $100 \pm \dots$  &  $0.135 \pm 0.014$  &  \dots \\
2012.38  &  56065.54 &  6  & $0.004 \pm 0.001$ & $1.78 \pm 0.06$  & $100 \pm \dots$  &  $0.100 \pm 0.010$  &  \dots \\
\noalign{\medskip}
2009.78  &  55117.82 &  2  & $0.031 \pm 0.003$ & $0.86 \pm 0.05$  & $103 \pm 1  $    &  $0.065 \pm 0.006$  &  C1    \\
2010.34  &  55323.28 &  3  & $0.008 \pm 0.001$ & $0.97 \pm 0.08$  & $93  \pm 5  $    &  $0.018 \pm 0.002$  &  C1    \\
2012.38  &  56065.54 &  6  & $0.010 \pm 0.001$ & $1.50 \pm 0.06$  & $107 \pm 1  $    &  $0.052 \pm 0.005$  &  C1    \\
\noalign{\medskip}
2009.35  &  54959.36 &  1  & $0.010 \pm 0.001$ & $0.49 \pm 0.06$  & $102 \pm 7  $    &  $0.016 \pm 0.002$  &  C2    \\
2009.78  &  55117.82 &  2  & $0.033 \pm 0.003$ & $0.46 \pm 0.05$  & $106 \pm 2  $    &  $0.061 \pm 0.006$  &  C2    \\
2010.34  &  55323.28 &  3  & $0.018 \pm 0.002$ & $0.62 \pm 0.05$  & $117 \pm 5  $    &  $0.269 \pm 0.027$  &  C2    \\
2011.35  &  55688.76 &  4  & $0.013 \pm 0.001$ & $0.63 \pm 0.07$  & $84  \pm 6  $    &  $0.331 \pm 0.033$  &  C2    \\
2011.77  &  55843.86 &  5  & $0.024 \pm 0.002$ & $0.79 \pm 0.04$  & $110 \pm 3  $    &  $0.211 \pm 0.021$  &  C2    \\
2012.38  &  56065.54 &  6  & $0.011 \pm 0.001$ & $0.75 \pm 0.04$  & $124 \pm 3  $    &  $0.197 \pm 0.020$  &  C2    \\
\noalign{\medskip}
2009.35  &  54959.36 &  1  & $0.446 \pm 0.045$ & $0.07 \pm 0.06$  & $137 \pm 43 $    &  $0.016 \pm 0.002$  &  C3    \\
2009.78  &  55117.82 &  2  & $0.056 \pm 0.006$ & $0.15 \pm 0.05$  & $110 \pm 19  $    &  $0.016 \pm 0.002$  &  C3    \\
2010.34  &  55323.28 &  3  & $0.014 \pm 0.001$ & $0.20 \pm 0.08$  & $132 \pm 22  $    &  $0.024 \pm 0.002$  &  C3    \\
2011.35  &  55688.76 &  4  & $0.023 \pm 0.002$ & $0.27 \pm 0.07$  & $101 \pm 15  $    &  $0.016 \pm 0.002$  &  C3    \\
2012.38  &  56065.54 &  6  & $0.009 \pm 0.001$ & $0.36 \pm 0.06$  & $142 \pm 10  $    &  $0.015 \pm 0.002$  &  C3    \\
\noalign{\medskip}
2009.35  &  54959.36 &  1  & $0.032 \pm 0.003$ & $0.25 \pm 0.06$  & $134 \pm 14  $    &  $0.016 \pm 0.002$  &  \dots \\
2009.78  &  55117.82 &  2  & $0.016 \pm 0.002$ & $0.27 \pm 0.05$  & $99  \pm 11  $    &  $0.016 \pm 0.002$  &  \dots \\
\noalign{\medskip}
2011.35  &  55688.76 &  4  & $0.032 \pm 0.003$ & $0.06 \pm 0.07$  & $30  \pm 5  $    &  $0.029 \pm 0.003$  &  \dots \\
2011.77  &  55843.86 &  5  & $0.024 \pm 0.002$ & $0.17 \pm 0.06$  & $103 \pm 19  $    &  $0.022 \pm 0.002$  &  \dots \\
\noalign{\medskip}
2009.35  &  54959.36 &  1  & $2.293 \pm 0.229$ & $0.00 \pm 0.06$  & $0.0   \pm 0.0  $   &  $0.031 \pm 0.003$  &  Core  \\
2009.78  &  55117.82 &  2  & $0.966 \pm 0.097$ & $0.00 \pm 0.05$  & $0.0   \pm 0.0  $   &  $0.029 \pm 0.003$  &  Core  \\
2010.34  &  55323.28 &  3  & $0.564 \pm 0.056$ & $0.00 \pm 0.08$  & $0.0   \pm 0.0  $   &  $0.038 \pm 0.004$  &  Core  \\
2011.35  &  55688.76 &  4  & $0.808 \pm 0.081$ & $0.00 \pm 0.07$  & $0.0   \pm 0.0  $   &  $0.026 \pm 0.003$  &  Core  \\
2011.77  &  55843.86 &  5  & $0.612 \pm 0.061$ & $0.00 \pm 0.06$  & $0.0   \pm 0.0  $   &  $0.036 \pm 0.004$  &  Core  \\
2012.38  &  56065.54 &  6  & $0.583 \pm 0.058$ & $0.00 \pm 0.06$  & $0.0   \pm 0.0  $   &  $0.027 \pm 0.003$  &  Core  \\
\noalign{\medskip}
\end{longtable}
\tablefoot{Columns from left to right:
(1) observing epoch in fractional year;
(2) MJD of the observing epoch;
(3) epoch identifier between $1$--$6$; 
(4) integrated component flux density;
(5) radial separation from the core;
(6) position angle;
(7) component size given as the FWHM of the major axis;
(8) component identification label.}
\end{longtab}

\begin{appendix}
\section{VLBI component light curves}\label{App:A}

\begin{figure*}
 \centering
 \includegraphics[width=\linewidth]{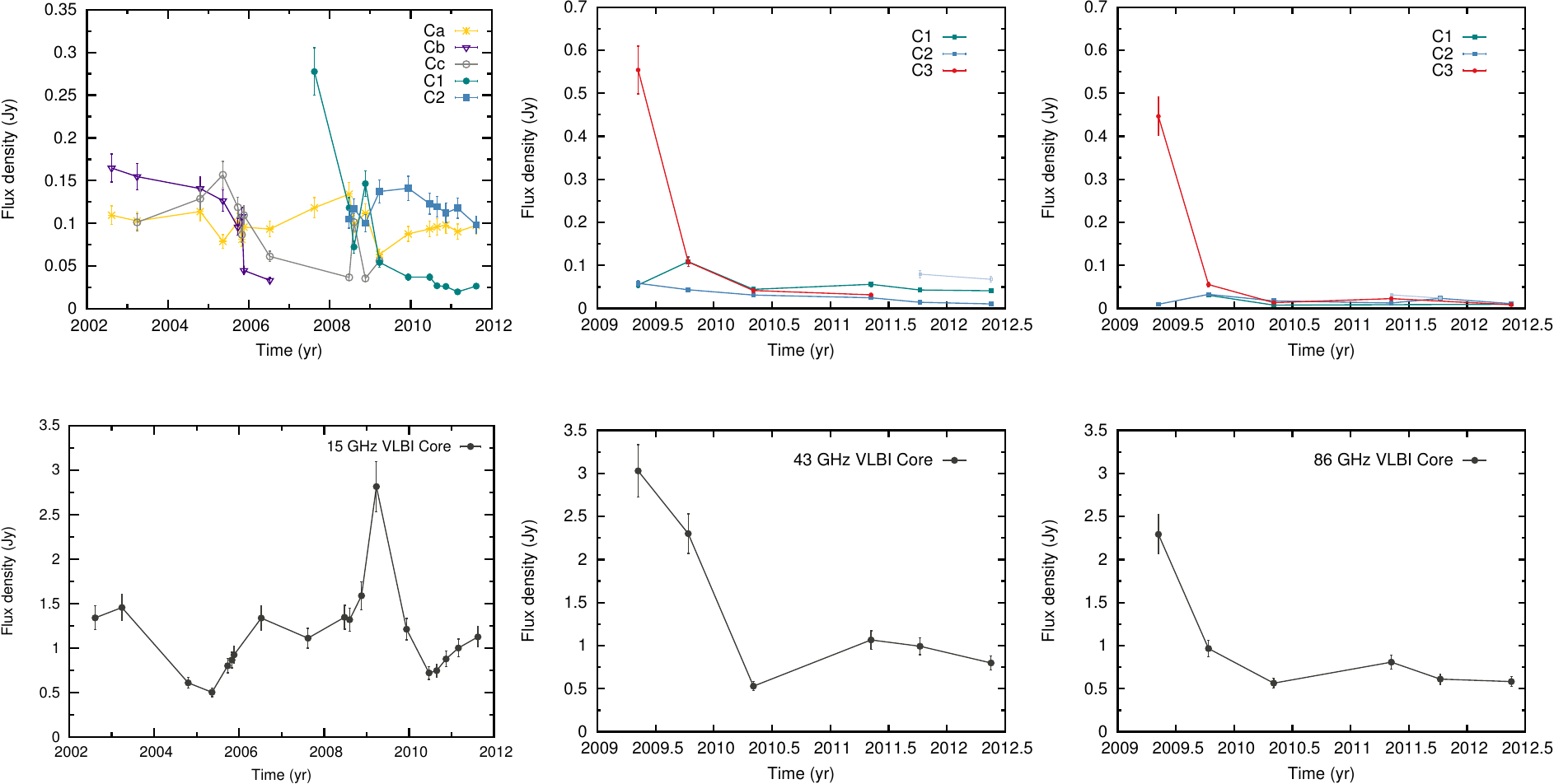}
 \caption{Light curves of individual components (top panels) and the core
(bottom panels) at $15$, 43\,GHz, and 86\,GHz. Component C1 shows significant
variability between $2008$ and $2010$ (at $15$\,GHz). Also note component C3 at
its decaying flux density phase, seen only at high-frequency observations.}
   \label{compLCsall}
\end{figure*}

The appendix plots feature all VLBI component light curves at all observing
frequencies (Fig. \ref{compLCsall}).
\end{appendix}

\begin{acknowledgements}
VK acknowledges the help of E. Ros during discussions and for his comments that
improved the manuscript.

Thanks also go to the anonymous referee. Her/his valuable comments increased the
quality of the paper.

VK was supported for this research through a stipend from the International
Max Planck Research School (IMPRS) for Astronomy and Astrophysics at
the Universities of Bonn and Cologne.

This research has made use of data from the MOJAVE database that is maintained
by the MOJAVE team.

This research has made use of data obtained with the Global Millimeter
VLBI Array (GMVA), which consists of telescopes operated by the MPIfR, IRAM,
Onsala, Metsahovi, Yebes, and the VLBA. The data were correlated at the
correlator of the MPIfR in Bonn, Germany. The VLBA is an instrument of the
National Radio Astronomy Observatory, a facility of the National Science
Foundation operated under cooperative agreement by Associated Universities, Inc.

Partly based on observations with the 100-m telescope of the MPIfR
(Max-Planck-Institut f\"{u}r Radioastronomie) at Effelsberg and
observations carried out with the IRAM 30-m Telescope. IRAM is
supported by INSU/CNRS (France), MPG (Germany) and IGN (Spain).

The single-dish millimetre observations were closely coordinated with the more
general flux density monitoring conducted by IRAM (Institut de Radioastronomie
Millimétrique), and data from both programs are included in this paper.
\end{acknowledgements}

\bibliographystyle{aa} 
\bibliography{Karamanavis_et_al_PKS1502+106_VLBI}

\end{document}